\documentclass[aps,twocolumn,prd,10pt,showpacs,showkeys,preprintnumbers,superscriptaddress,nobibnotes,nofootinbib,floatfix,longbibliography]{revtex4-1}

\usepackage{graphicx}
\usepackage{dcolumn}% Align table columns on decimal point
\usepackage{bm}% bold math
\usepackage{hyperref}
\usepackage[utf8]{inputenc}
\usepackage[dvipsnames]{xcolor}
\usepackage[utf8]{inputenc}
\usepackage{xspace}

\graphicspath{{figures/}{supplementary_figures/}}

\newcommand{\geant}{\textsc{Geant4}\xspace}
\newcommand{\genie}{\textsc{genie}\xspace}
\newcommand{\gibuu}{\textsc{g}i\textsc{buu}\xspace}

\begin{document}

\title{Lepton-Nucleus Cross Section Measurements for DUNE with the LDMX Detector}

\author{Artur~M.~Ankowski}
\affiliation{SLAC National Accelerator Laboratory, Menlo Park, CA 94025, USA}

\author{Alexander~Friedland}
\affiliation{SLAC National Accelerator Laboratory, Menlo Park, CA 94025, USA}

\author{Shirley~Weishi~Li}
\affiliation{SLAC National Accelerator Laboratory, Menlo Park, CA 94025, USA}

\author{Omar~Moreno}
\affiliation{SLAC National Accelerator Laboratory, Menlo Park, CA 94025, USA}

\author{Philip~Schuster}
\affiliation{SLAC National Accelerator Laboratory, Menlo Park, CA 94025, USA}

\author{Natalia~Toro}
\affiliation{SLAC National Accelerator Laboratory, Menlo Park, CA 94025, USA}

\author{Nhan~Tran}
\affiliation{Fermi National Accelerator Laboratory, Batavia, IL 60510, USA}

\date{April 5, 2020}

\begin{abstract}
We point out that the LDMX (Light Dark Matter eXperiment) detector design, conceived to search for sub-GeV dark matter, will also have very advantageous characteristics to pursue electron-nucleus scattering measurements of direct relevance to the neutrino program at DUNE and elsewhere. These characteristics include a 4-GeV electron beam, a precision tracker, electromagnetic and hadronic calorimeters with near 2$\pi$ azimuthal acceptance from the forward beam axis out to $\sim$40$^\circ$ angle, and low reconstruction energy threshold. LDMX thus could provide (semi)exclusive cross section measurements, with detailed information about final-state electrons, pions, protons, and neutrons. We compare the predictions of two widely used neutrino generators (\genie, \gibuu) in the LDMX region of acceptance to illustrate the large modeling discrepancies in electron-nucleus interactions at DUNE-like kinematics. We argue that discriminating between these predictions is well within the capabilities of the LDMX detector.
\end{abstract}

\preprint{SLAC-PUB-17494, FERMILAB-PUB-19-619-SCD}

\maketitle

%%%%%%%%%%%%%%%%%%%%%%%%%%%%%%%%%%%%%%%%%%%%%%%%%%%%%%%%%%%%%%
%%%%%%%%%%%%%%%%%%%%%%%%%%%%%%%%%%%%%%%%%%%%%%%%%%%%%%%%%%%%%%

\section{Introduction}

The discovery of neutrino masses and flavor mixing represents a breakthrough in the search for physics beyond the Standard Model.  As the field of neutrino physics enters the precision era, accelerator-based neutrino oscillation experiments are taking center stage.  This includes NOvA, T2K, and MicroBooNE, which are currently taking data, SBND and ICARUS detectors, which will soon be deployed at Fermilab, and the Deep Underground Neutrino Experiment (DUNE), for which the technical design is being finalized.

%%%%%%%%%%%%%%%%%%%%%%%%%
\begin{figure}
    \begin{center}
        \includegraphics[width=\columnwidth]{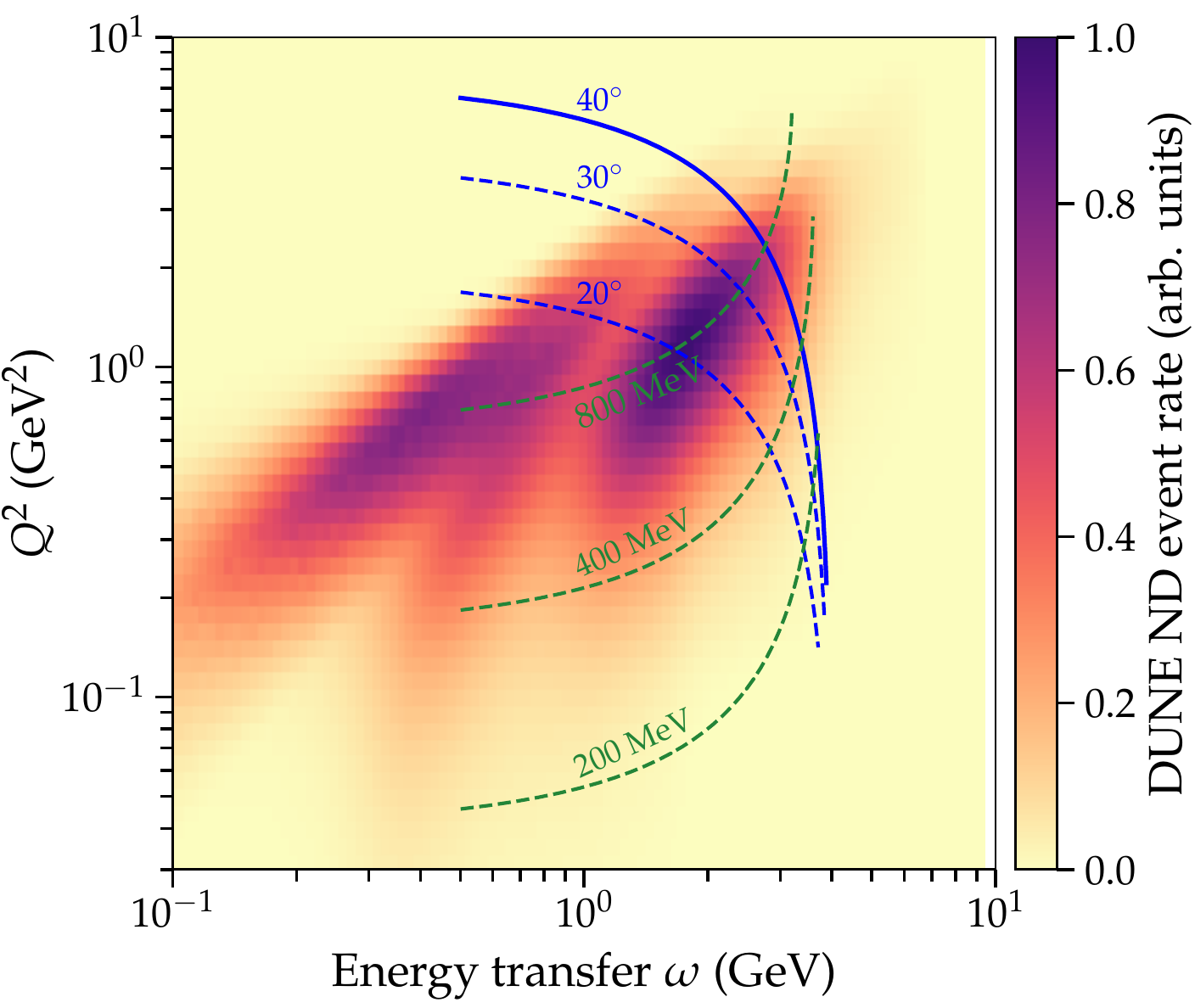}
       \caption{Simulated event distribution for charged-current muon neutrino scattering on argon in the DUNE near detector, shown as a heat map, compared with the kinematics accessible in \emph{inclusive and (semi)exclusive} electron scattering measurements at LDMX. Blue lines correspond to constant electron-scattering angles of 40$^\circ$, 30$^\circ$, and 20$^\circ$. Green lines represent contours of constant transverse electron momenta $p_T$ of 800, 400, and 200 MeV.  As currently envisioned, LDMX can probe the region with $\theta_e <40^\circ$ and $p_T >10$~MeV (below the scale of the plot).}
        \label{fig:DUNE_distribution}
    \end{center}
\end{figure}
%%%%%%%%%%%%%%%%%%%%%%%%%

The primary goal of the accelerator-based neutrino program is the measurement of oscillation features in a reconstructed neutrino-energy spectrum.  Performing this reconstruction accurately and consistently for both neutrinos and antineutrinos requires a detailed understanding of how (anti)neutrinos interact with nuclei---a subtlety that has already impacted past oscillation fits~\cite{Adamson:2016xxw, Adamson:2017qqn, NOvA:2018gge},  despite the availability of near detectors, which can help tune cross section models and constrain other systematic effects.  The situation will be even more challenging at DUNE~\cite{Ankowski:2015kya}, where the science goal is to measure the subtle effects of $\delta_{\text{CP}}$ and mass hierarchy, requiring a much higher level of precision.

The origin of these difficulties stems from the complexity of neutrino-nucleus interactions in the relevant energy range, which for DUNE is approximately between 500~MeV and 4~GeV. At these energies, different mechanisms of interaction yield comparable contributions to the cross section (see Appendix~\ref{sec:DUNE_distri} for details).  One has to model both quasielastic (QE) scattering, in which a struck nucleon remains unbroken, $\nu_\mu +n\rightarrow \mu^{-} +p$, and various processes in which one or more pions are produced. The latter can occur through the excitation of baryonic resonances, as well as through nonresonant channels. At sufficiently high values of four-momentum transfer, $Q^2=-(p_{\nu}-p_{\mu})^{2}$, and energy transfer, $\omega= E_\nu - E_\mu$, the deep inelastic scattering (DIS) description of the interaction becomes appropriate, in which the lepton scatters on individual quarks inside the nucleon, followed by a process of ``hadronization.''

As DUNE uses argon as a target, all this happens inside a large nucleus, adding further complexity. The presence of the surrounding nucleons means hadrons created at the primary interaction vertex may undergo large final-state interactions (FSI) on their way out. The resulting intranuclear cascade can lead to energy loss or absorption of primary hadrons, production of additional hadrons, and nucleon knockout. Initial states of the interacting nucleons are also affected, by nuclear binding and motions inside a nucleus. Last but not least, multinucleon effects, such as meson-exchange currents (MEC), which arise from scattering on interacting nucleon pairs, likewise have to be considered.

To model this rich physics, experiments rely on event generator codes, among them \genie~\cite{Andreopoulos:2009rq,Andreopoulos:2015wxa} and \gibuu~\cite{Buss:2011mx,Gallmeister:2016dnq, Mosel:2017nzk, Mosel:2017ssx, Mosel:2019vhx}, which are used as benchmarks in this paper. As we will see explicitly below, these codes are often not in agreement with each other. More importantly, they are often also not in agreement with recent high-statistics data from the MINERvA experiment, collected in the kinematic regime relevant to DUNE. For example, the default models in \genie seem to significantly overestimate neutron production~\cite{Elkins:2019vmy}, mispredict the ratio of charge-current interactions across different nuclear targets~\cite{Tice:2014pgu}, and mismodel single-pion production~\cite{Stowell:2019zsh}. Thus, there is direct \emph{experimental} evidence that existing models need to be improved.

Importantly, simple phenomenological tuning of parameters within the existing models may not be sufficient. For example, Ref.~\cite{Stowell:2019zsh} reports that no tune could describe all different exclusive final states in their analysis. Crucially, the paper also notes that the physical origin of the discrepancies is difficult to pinpoint, based on only the available data.

This brings us to an important question: what new data are needed to improve the physics in these generators?  A priori, one might think that all that is needed is more neutrino-nucleus scattering data, with higher statistics and precision, as will be collected with the future near detectors. In reality, while better neutrino data would certainly be desirable, it is unlikely to be sufficient. To date, neutrino experiments only have access to broadband beams, extract flux-integrated cross sections~\cite{AguilarArevalo:2010zc,Rodrigues:2015hik,Patrick:2018gvi,Gran:2018fxa,Ruterbories:2018gub,Carneiro:2019jds,Adams:2019iqc,Abe:2019sah,Abe:2018uhf}, and neutrino-energy reconstruction itself suffers from sizable uncertainties. In turn, the process of energy reconstruction \emph{relies} on neutrino generators. The reason is that even today's state-of-the-art neutrino detectors are imperfect calorimeters at several GeV energies, with event generators being used to fill in the missing information. Hence, complementary probes that are free from these limitations are highly desirable for accurately validating the physical models in event generators. 

Precise electron-nucleus scattering data provide just such a complementary probe. While electron and neutrino interactions are different at the primary vertex, many relevant physical processes in the nucleus are the same in the two cases, as discussed below in Sec.~\ref{sec:e-scattering}. What electron scattering offers is precisely controlled kinematics (initial and final energies and scattering angles), large statistics, \emph{in situ} calibration of the detector response using exclusive reactions, and a prospect of easily swapping different nuclear targets. This allows one to easily zero in on specific scattering processes and to diagnose problems that are currently obscured by the quality of the neutrino scattering data. 

In this paper, we point out that the proposed LDMX (Light Dark Matter eXperiment) setup at SLAC~\cite{Akesson:2018vlm}, designed to search for sub-GeV dark matter, will have very advantageous characteristics to also pursue electron-scattering measurements relevant to the neutrino program. These include a 4-GeV electron beam and a detector with high acceptance of hadronic products in the $\sim$40$^\circ$ forward cone and low-energy threshold. Figure~\ref{fig:DUNE_distribution} shows the distribution, in the $(\omega,Q^2)$ plane, of charged-current (CC) events for muon neutrino scattering on argon nuclei in the near detector of DUNE, simulated with the \gibuu generator code. As can be immediately seen, the LDMX coverage in the relevant kinematic window is excellent. Below, we quantify how future LDMX data can be used to test and improve physics models in lepton-nucleus event generator codes.

%%%%%%%%%%%%%%%%%%%%%%%%%%%%%%%%%%%%%%%%%%%%%%%%%%%%%%%%%%%%%%%%%%%%%%%%%%%%%%
%%%%%%%%%%%%%%%%%%%%%%%%%%%%%%%%%%%%%%%%%%%%%%%%%%%%%%%%%%%%%%%%%%%%%%%%%%%%%%

\section{Electron scattering measurements and neutrino cross sections}
\label{sec:e-scattering}

Let us now define the connection between electron- and neutrino-nucleus scattering more precisely. Superficially, the mere existence of such a connection is not obvious, since the weak and electromagnetic forces have a number of important differences. The differences are immediately apparent in the elastic scattering regime: while CC neutrino interactions occur on initial-state neutrons in the nucleus, electromagnetic scattering also involves initial-state protons (neutrons couple through their magnetic moments). The situation is similar in the DIS regime, where the primary vertex is treated at the quark level: while CC neutrino (antineutrino) interactions are controlled by the distribution of initial-state down (up) quarks, electron scattering involves both up and down quarks. Additional differences come from the chiral nature of the weak interactions. While the electron-nucleon vertex is sensitive only to the electric charge distribution inside a nucleon and its magnetic moment, neutrino scattering also depends on the distribution of the axial charge. The effect of this axial coupling is not small; in fact, at 1-GeV neutrino energy, the axial part of the weak interaction provides a dominant contribution to the elastic neutrino-nucleus cross section. In short, one should not expect to blindly convert electron-scattering data into predictions for neutrinos.

Yet, a tight connection between electron and neutrino scattering does exist. This is most immediately seen by considering the nuclear physics of the problem. Neutrino scattering depends on the wave functions of the initial nucleons (in momentum space) and on the nuclear density profile, and these are most accurately probed with electron scattering.  The differences between proton and neutron couplings mentioned above might give one pause. However, by systematically analyzing electron data on mirror nuclei, in which neutrons in one nucleus have the same shell structure as protons in another, one can learn about both proton and neutron wave functions~\cite{Bohr,Cohen,Dai:2018gch}. 

The same argument can be made about modeling final-state interactions, which dictate the subsequent evolution of the interaction products inside the nucleus~\cite{Glauber:1970jm,Horikawa:1980cg,Buss:2011mx}. FSI can significantly modify the properties of the hadronic system, through energy loss of propagating particles, absorption and creation of mesons, as well as nucleon knockout. It is essential to model the intranuclear transport of various hadrons using a unified framework, regardless of whether they are produced in electron or neutrino scattering. The accuracy of the treatment can then be validated by targeted studies of exclusive hadronic final states in electron scattering.

We see here that it is important for the differences between electron and neutrino interactions to be limited to the elementary scattering vertex. This is justified at typical momentum-transfer values relevant to DUNE, where scattering involves predominantly a single nucleon. It can be shown that, under these conditions, nuclear effects become largely independent of the interaction dynamics~\cite{Benhar:2005dj}.

The connections between electron and neutrino scattering, in fact, extend beyond nuclear physics models, to include many hadronic physics effects. For example, to model neutrino-quark interactions in the DIS regime, one needs accurate parton distribution functions. These can be extracted from precision electron-scattering data. The physics of the subsequent hadronization can also be treated in a common framework. Finally, it is desirable to use a unified treatment of other physics, such as hadronic resonances, two-nucleon currents, or quark-hadron duality. Of course, in doing so, one needs to include the correct treatment of the nucleon axial properties. Even there, however, comparisons to electron scattering are proving to be highly advantageous. For example, recent lattice QCD studies found it useful to simultaneously model the nucleon axial and vector form factors (see, e.g., Refs.~\cite{Capitani:2015sba,Capitani:2017qpc,Alexandrou:2018pln,Jang:2019vkm,Jang:2019jkn}).

The importance of using the same nuclear model for neutrino and electron scattering was realized a long time ago, as illustrated, for instance, by the discussion in the seminal paper by Smith and Moniz~\cite{Smith:1972xh}. In fact, it was argued in that paper that combining electron and neutrino scattering gives one the best tool for probing the physics of the nucleus. The same argument has also been made more recently from the experimental point of view~\cite{Gallagher:2004nq}. It has since been incorporated into the mission statement of the GENIE generator. Insofar as this crucial principle is adhered to in the generator development and applications, electron-scattering data should provide an excellent validation platform.

Let us next outline the requirements from the point of view of neutrino experiments. As stated in the Introduction, the key to many modern neutrino experiments is accurate neutrino-energy reconstruction. Experiments such as NOvA and DUNE approach this problem by using the calorimetric technique, which involves adding up visible energies of all final-state particles and inferring invisible components, such as neutrons and low-energy charged hadrons, using event-generator predictions. Event generators are also used to model the composition of the final-state hadronic system, whenever that information is unavailable from the particle-identification algorithms. Knowledge of the final-state composition is needed to convert measured ionization charge, or scintillation light, to true energy loss. This is not a small effect, and existing differences among generator models consistent with available validation data can yield energy reconstruction variations as large as 20\%, which has been discussed systematically in Ref.~\cite{Friedland:2018vry}, together with other factors impacting the energy resolution.

Thus, to adequately constrain the underlying generator models, one needs to measure not only inclusive electron-scattering rates, but also collect detailed information about the exclusive hadronic final states. This includes charged pions, neutral pions, and protons, as well as any available information on final-state neutrons. Practically, one needs to simultaneously measure the kinematics of an energetic, often-forward electron, as well as detect charged hadrons to below 100--200~MeV momenta (see, e.g., Ref.~\cite{Eberly:2014mra}) with wide and well-characterized angular acceptance.

Discrepancies between scattering data and generator predictions can indicate problems either with the nuclear model or with hadronic physics~\cite{AFtoappear,GrandUnifiedtoappear}. Having information on exclusive hadronic final states can help diagnose the origin of the problem. To conclusively disentangle nuclear and hadronic effects may require comparative analyses of electron-scattering data on various nuclear targets, including the lightest elements---helium, deuterium, and hydrogen. That such targets can be quite small in the case of electron scattering represents another tangible advantage over neutrino scattering, where concerns about fire safety make future hydrogen bubble-chamber experiments prohibitively costly.

To this end, a systematic analysis of the data collected on various nuclear targets by different experiments using the CLAS detector in Hall B at Jefferson Laboratory, while not completely addressing the requirements outlined above, would be an important advance. So far, the published studies focused on specific hadronic processes with hydrogen targets~\cite{Ripani:2002ss,Fedotov:2008aa,Bedlinskiy:2014tvi,Park:2014yea,Isupov:2017lnd,Mattione:2017fxc,Fedotov:2018oan,Markov:2019fjy}.  These should already be useful for testing generator models for certain hadronic processes, such as $\rho$ meson production through higher resonances. The CLAS12 proposal ``Electrons for Neutrinos'' would make further inroads by collecting more data~\cite{e4nu17,e4nu18}. At present, published datasets involving argon and its mirror nucleus titanium come from a separate experiment in Hall A~\cite{Benhar:2014nca,Dai:2018gch,Dai:2018xhi}. While undoubtedly valuable~\cite{Bodek:2018lmc,Mosel:2018qmv,Barbaro:2019vsr,Barbieri:2019ual,Gonzalez-Jimenez:2019ejf}---for example, enabling comparisons with the well-studied carbon data~\cite{AFtoappear}---they are limited to the inclusive spectrum of scattered electrons measured at a single value of the beam energy (2.22 GeV) and a fixed scattering angle ($15.54^\circ$).

At the moment, and over the next several years, electronuclear scattering data with excellent hadronic final-state reconstruction is sorely needed.  The ideal would be reconstruction with no detection threshold, full $4\pi$ coverage, and with excellent neutron identification. While CLAS12 can make some inroads in this direction, its acceptance will be limited (especially in the forward direction) and neutron-energy reconstruction will be modest. The proposed LDMX detector concept offers a number of complementary and unique advantages that can be leveraged to provide a range of valuable electron-nucleus scattering data for the purpose of constraining neutrino-scattering models.

%%%%%%%%%%%%%%%%%%%%%%%%%%%%%%%%%%%%%%%%%%%%%%%%%%%%%%%%%%%%%%%%%%%%%%%%%%%%%%
%%%%%%%%%%%%%%%%%%%%%%%%%%%%%%%%%%%%%%%%%%%%%%%%%%%%%%%%%%%%%%%%%%%%%%%%%%%%%%

\section{The LDMX Detector Concept and Electron-Nucleus Scattering Data}
\label{sec:LDMX}

%%%%%%%%%%%%%%%%%%%%%%%%%
\begin{figure}[t]
	\begin{center}
        \includegraphics[width=0.9\columnwidth]{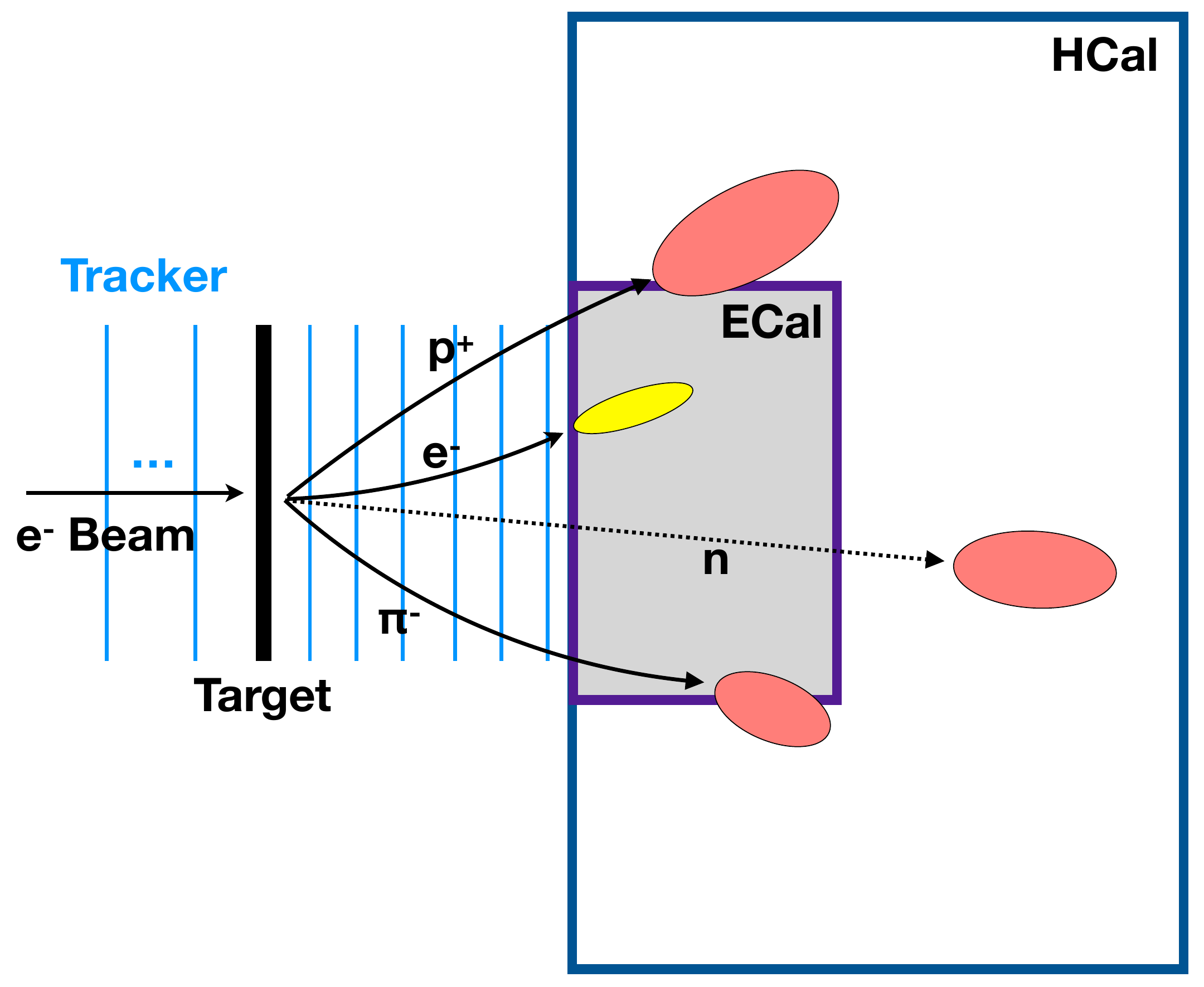}
        \caption{Schematic of the LDMX experiment for dark-matter search (not to scale).  The electron beam is incident from the left and interacts in the target (which can be varied). Direct tracking and calorimetry along the beam axis provides excellent (nearly $2\pi$ azimuthal) forward acceptance to a range of final-state particles, including the recoiling electron, protons, pions, and neutrons.}
        \label{fig:ldmx}
    \end{center}
\end{figure}
%%%%%%%%%%%%%%%%%%%%%%%%%

LDMX is a fixed-target experiment designed to search for sub-GeV dark matter, employing a high-repetition rate, low-current electron beam~\cite{Akesson:2018vlm} with precision tracking (in a magnetic field) and calorimetry along the beam axis to provide high-fidelity detection of both charged and neutral particles. Figure~\ref{fig:ldmx} provides a high-level illustration of the detector layout, which is largely optimized to search for dark-matter production. In candidate events for dark-matter production, most of the initial electron's energy is expected to be carried away by undetected particle(s). Therefore, identification of these processes requires an excellent hermeticity of the detector, allowing, e.g., energetic neutron-knockout events to be detected with sufficiently small uncertainty. In fact, the primary purpose of the downstream calori{\-}metry in LDMX is to provide a fast, radiation-hard, and highly granular veto against photonuclear and electronuclear reactions in the target area that might generate difficult-to-detect final states, and hence a potential background to dark-matter reactions. In the nominal design, the vast majority of triggered data would be composed of these photo/electronuclear reactions, and rejected offline. The key result of this paper is that this vetoed data will itself be of great value in service of neutrino-interaction modeling, as was described above. 

To see why this is the case, we start with a more detailed description of the detector layout. The tracking system upstream of the target and the target itself are housed inside of a 1.5-T dipole magnet while the downstream (recoil) tracker is in the fringe magnetic field. The target is currently envisioned to be titanium, and we assume it to be $0.1~X_0$ (0.356 cm) thick, $X_0$ being the radiation length. However, different target materials (such as argon) and thicknesses are possible, as discussed further in Sec.~\ref{sec:future}. The two tracking systems provide robust measurements of incoming and outgoing electron momentum. 

The Electromagnetic Calorimeter (ECal) is surround{\-}ed by the Hadronic Calorimeter (HCal) to provide large angular coverage downstream of the target area, in order to efficiently detect interaction products.  The ECal is a silicon-tungsten high-granularity sampling calorimeter based on a similar detector developed for the high-luminosity Large Hadron Collider upgrade of the endcap calorimeter of the Compact Muon Solenoid (CMS) detector. The ECal is radiation tolerant with fast readout, and the high granularity provides good energy resolution and shower discrimination for electromagnetic and hadronic interactions.  The HCal is a scintillator-steel sampling calorimeter that has wide angular coverage and is sufficiently deep to provide required high efficiency for detecting minimum ionizing particles and neutral hadrons.

While the final detector design is still under development, we describe a coarse set of detector capabilities (motivated by the baseline design), which are particularly relevant for electron-scattering measurements~\cite{Akesson:2018vlm}: 
%%%%%%%%%%%%%%%%%%%%%%%%% 
\begin{itemize}
    \item \emph{Electrons}: We estimate the electron energy resolution to be 5\%--10\% and the $p_T$ resolution to be $<10$~MeV~\cite{Akesson:2018vlm}, where $p_T$ is the transverse momentum of the outgoing electron.  The tracker acceptance is approximately 40$^\circ$ in the polar angle where the $z$-axis is defined along the beamline. Electrons can be measured down to a kinetic energy of approximately 60~MeV.
    \item \emph{Charged pions and protons}: The energy and $p_T$ resolutions, tracking acceptance, and kinetic thresholds are similar for charged pions, protons, and electrons.  The estimate of tracking angular and momentum acceptance is shown in Fig.~\ref{fig:tracking}.  The recoil tracker and ECal detectors can be used to perform particle identification via mean energy loss ($dE/dx$) to separate charged pions and protons.  Based on previous studies of similar silicon-tracking technologies at CMS~\cite{Khachatryan:2010pw, Giammanco:2011zz}, the recoil tracker by itself has good pion/proton discrimination power for kinetic energies $<1.5$~GeV.
    \item \emph{Neutrons}: The nominal neutron signal is a hadronic shower in the HCal, although the shower can start in the ECal, which is roughly one hadronic-interaction mean free path in thickness.  The signature also requires that there be no charged (minimum ionizing particle) track aligned with the shower in the tracking (ECal) system.  Identifying/reconstructing single neutrons will rely on localized and separable hadronic showers.  Once identified, neutrons can be efficiently distinguished from charged hadrons (protons, charged pions/kaons) at angles $<40^\circ$ by identifying those charged tracks in the tracking and ECal detectors.
    
    Based on \geant simulations for the baseline HCal sampling fraction, we estimate the HCal to have an energy resolution for neutrons of $5\% \oplus 40\%/\sqrt{E/\text{GeV}}$ and a polar angular acceptance of 65$^\circ$.  However, because we have tracking acceptance out to $\sim$40$^\circ$, our studies assume that we have good pion/proton/neutron discrimination out to only $\sim$40$^\circ$~\cite{Akesson:2018vlm}.  We have also assumed that the angular resolution of the neutrons are conservatively 10$^\circ$ based on position resolution measurements.  
    We leave it to future studies to understand additional separation power between 40$^\circ$ and 65$^\circ$. 
    
    Of course, detecting a detached cluster in the HCal does not guarantee that it was created by a neutron that came from the primary electron interaction vertex. Some neutrons can be created in secondary and tertiary interactions of energetic charged hadrons. Understanding how well the primary neutron component can be isolated requires dedicated future simulations.
    
%%%%%%%%%%%%%%%%%%%%%%%%%
\begin{figure}
	\begin{center}
        \includegraphics[width=\columnwidth]{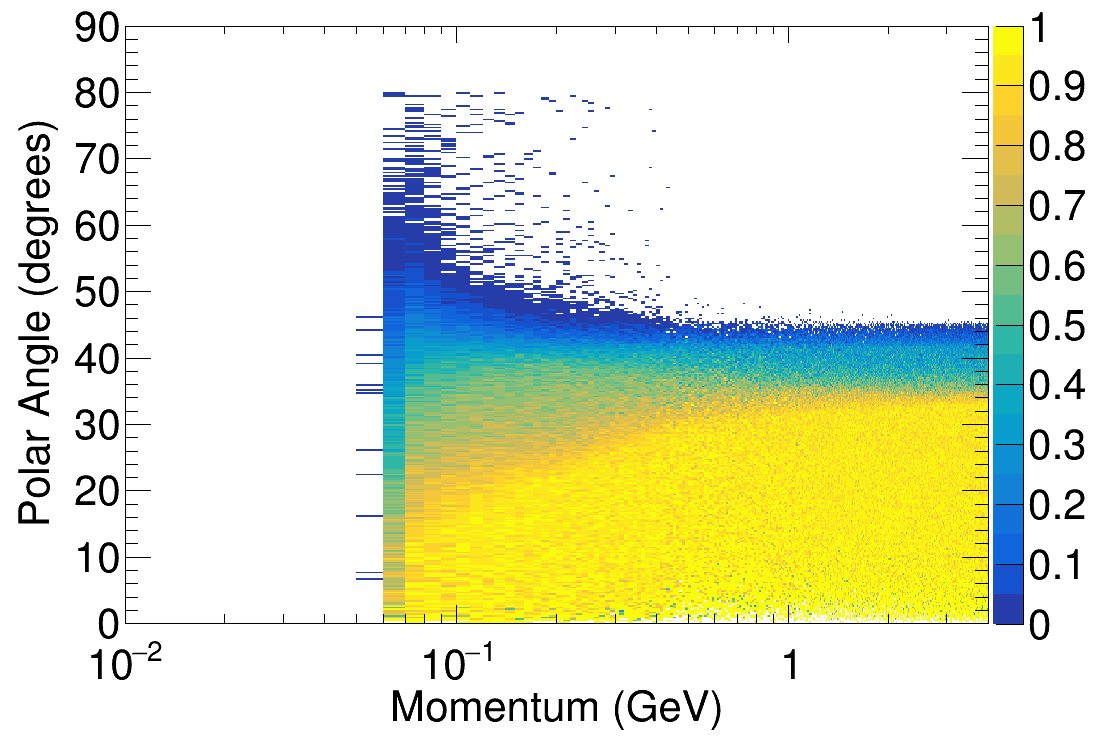}
        \caption{The acceptance of a charged particle (pion) track as a function of its momentum and polar angle.  The acceptance is defined as a charged particle that leaves four hits in the recoil tracking system.}
        \label{fig:tracking}
    \end{center}
\end{figure}
%%%%%%%%%%%%%%%%%%%%%%%%%

    \item \emph{Readout rate}: The total data acquisition (DAQ) rate of the detector is approximately 5~kHz.  A significant fraction of the DAQ bandwidth targets high-energy-transfer reactions. Thus, for this study, we focus on electron energy transfer $\omega > 1$~GeV.  This energy-transfer threshold is still below the nominal threshold for the dark-matter search, but could be achieved by prescaling the trigger or by using a combination of ECal and HCal online selections. Even smaller values of $\omega$ may be possible, but we leave such studies to future work. 
\end{itemize}
%%%%%%%%%%%%%%%%%%%%%%%%% 

For the studies described below, we assume a 4-GeV incoming electron beam and a dataset of $1\times 10^{14}$~EoT (electrons on target), corresponding to approximately 6 months of data collecting during an envisioned first phase of low-luminosity running.  The beam repetition rate is assumed to be 46~MHz and the beam is tuned to have  on average one electron per bucket.

With the beam and detector configurations described above, we will next explore the potential for LDMX to perform measurements of both inclusive (Sec.~\ref{sec:incl}) and (semi)exclusive (Sec.~\ref{sec:excl}) electron-nucleus scattering processes.  

%%%%%%%%%%%%%%%%%%%%%%%%%%%%%%%%%%%%%%%%%%%%%%%%%%%%%%%%%%%%%%%%%%%%%%%%%%%%%%
%%%%%%%%%%%%%%%%%%%%%%%%%%%%%%%%%%%%%%%%%%%%%%%%%%%%%%%%%%%%%%%%%%%%%%%%%%%%%%

\section{Monte Carlo Generators}
\label{sec:generators}

We study the modeling of electron-titanium interactions using the Monte Carlo generators \genie (versions 2.12.8 and 3.0.6)~\cite{Andreopoulos:2009rq, Andreopoulos:2015wxa} and \gibuu (versions 2017 and 2019)~\cite{Buss:2011mx,Mosel:2019vhx}. As both \genie and \gibuu had major updates, we show results obtained using both the versions before and after these changes. In the context of the inclusive cross sections, we also present the results obtained using \geant (version 4.10.p3)~\cite{Agostinelli:2002hh}, for reference. 

\genie~\cite{Andreopoulos:2009rq, Andreopoulos:2015wxa} is the generator most widely used in neutrino experiments and the default code employed in DUNE studies. In this analysis, we use its default configurations (``{\tt DefaultPlusMECWithNC}" for version 2.12 and ``{\tt EMPlusMEC\_G18\_02a\_00\_000}" for version 3.0). Nuclear effects are described in \genie within the global relativistic Fermi gas model of Bodek and Ritchie~\cite{Bodek:1980ar}. This approach treats the nucleus as a fragment of noninteracting nuclear matter of constant density, bound in a constant potential. The effect of short-range correlations between nucleons is added in an {\it ad hoc} manner, by extending the step-function momentum distribution above the Fermi momentum, $p_F\simeq240$--250 MeV, with a high-momentum tail. The binding energy is taken to be independent of momentum, and fixed to a value $\sim$30 MeV. As a consequence, nucleons in the high-momentum tail of the Bodek-Ritchie model are typically unbound. 

Pion production through excitation of nucleon resonances is described in \genie 2.12 using the framework of the Rein-Sehgal model~\cite{Rein:1980wg}.  While the original work included 18 resonances and accounted for interference between them, its implementation in \genie disregards the effect of interference, and is limited to 16 resonances, which are described using up-to-date parameters.  In \genie 3.0, the default model for resonance excitation is the approach of Berger and Sehgal~\cite{Berger:2008xs}.

All mechanisms of pion production on nucleons that do not involve resonance excitation are referred to in \genie as DIS processes. They are modeled following the effective approach of Bodek and Yang~\cite{Bodek:2002ps,Bodek:2004pc}. Relying on leading-order parton-distribution functions~\cite{Gluck:1998xa}, this model applies higher-order corrections to the effective masses of the target and the final state, in order to extend the applicability of the parton model to the low-$Q^2$ region. While DIS is the only mechanism of interaction in \genie for the invariant hadronic masses $W\geq1.7$ GeV, it is also employed to produce nonresonant background of events involving one or two pions in the resonance region, corresponding to $W<1.7$ GeV.

\gibuu~\cite{Buss:2011mx,Mosel:2019vhx} is a Monte Carlo code based on transport theory, originally developed to describe heavy ion collisions. Its nuclear model accounts for the nuclear density profile determined in electron scattering according to Ref.~\cite{DeJager:1974liz}, treating the nucleus as a local relativistic Fermi gas, bound by a potential exhibiting momentum dependence~\cite{Mosel:2018qmv}.

The implementation of both resonance-excitation processes and single-pion nonresonant background in \gibuu makes use of the MAID analysis~\cite{Drechsel:2007if}. MAID includes 13 resonances with invariant mass $W\leq2.0$ GeV and accounts for the interference between them, as well as for the interference between the resonant and nonresonant contributions. The two-pion rate is estimated by generalizing the model in Ref.~\cite{Nacher:2000eq} for photoproduction, by using the assumptions of Ref.~\cite{Christy:2007ve}.

To describe DIS processes, \gibuu relies on a modification of the {\sc pythia} code~\cite{Sjostrand:2006za}, extending its applicability down to the invariant hadronic mass 2.0 GeV. In this manner, leading order processes are implemented in the primary interaction vertex.

%%%%%%%%%%%%%%%%%%%%%%%%%
\begin{figure*}[!htb]
    \begin{center}
        \includegraphics[width=0.49\textwidth]{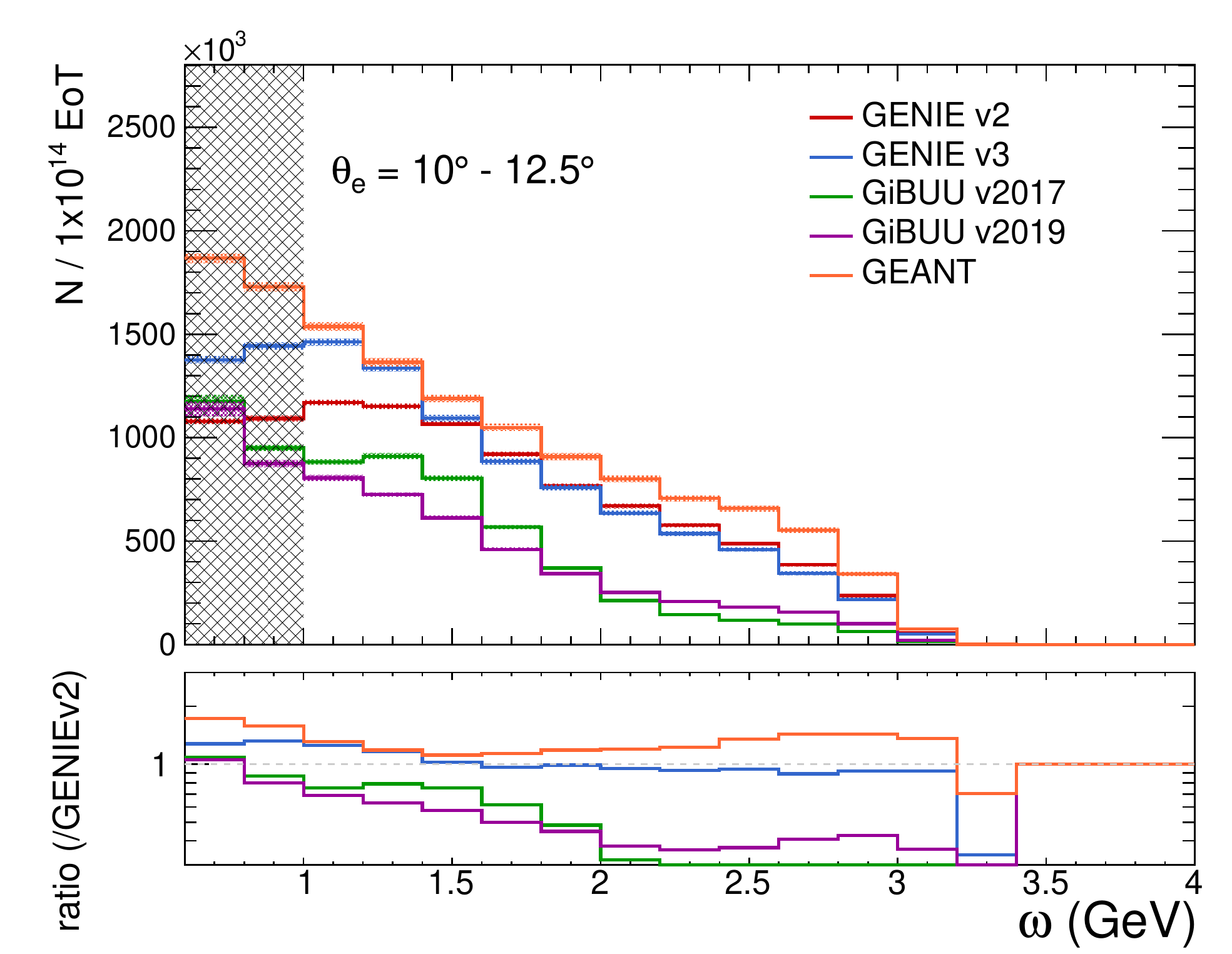}
        \includegraphics[width=0.49\textwidth]{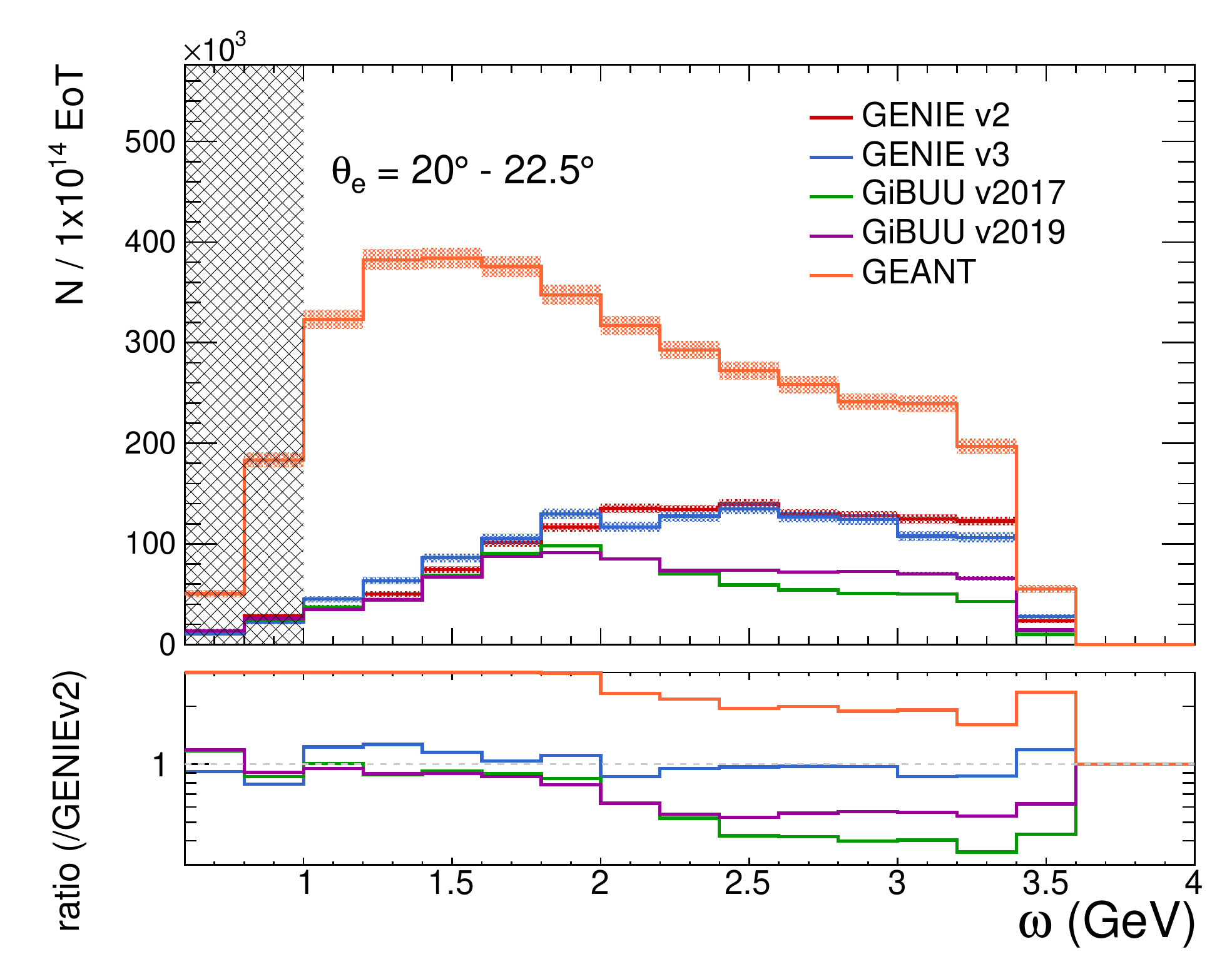}
        \caption{Event distribution as a function of electron energy transfer for the scattering angles of $10^{\circ}\leq\theta_e\leq12.5^{\circ}$ (left panel) and $20^{\circ}\leq\theta_e\leq22.5^{\circ}$ (right panel).  Scattered electrons are required to have the transverse momentum $p_T > 200$~MeV. In our nominal analyses, the trigger selection $\omega>1$ GeV is employed.}
        \label{fig:elomega}
    \end{center}
\end{figure*}
%%%%%%%%%%%%%%%%%%%%%%%%%

Performing simulations using the \geant generator~\cite{Agostinelli:2002hh}, we rely on the description of electron-nucleus interactions within the Bertini cascade model~\cite{Wright:2015xia} with improvements discussed in Ref.~\cite{Akesson:2018vlm}. This model relies on a parametrization of photoproduction data~\cite{Alekhin:1987nn} to obtain the elementary cross sections for electrons by employing the equivalent-photon approximation~\cite{vonWeizsacker:1934nji,Williams:1934ad}. The nuclear model of \geant, based on a local nonrelativistic Fermi gas model, approximates the density profile of a medium-sized nucleus as three regions of constant density~\cite{Wright:2015xia}. Within every region, the binding energy of nucleons depends on the local Fermi momentum and on the atomic charge and mass numbers. For pions, a universal constant potential is used.

In order to eliminate trivial differences between the three Monte Carlo generators, we apply the kinematic selection $Q^2 > 0.03~{\rm GeV}^2$~\cite{Dytman_private}, needed to define a phase space where all the generators are physically valid\footnote{As \genie generates events relying on $Q^2$ and the cross section for electrons is divergent when $Q^2\rightarrow0$, it is necessary to impose a cut on the minimal $Q^2$ value.}. This selection has no visible effect on the presented cross sections, required to pass our trigger selection, $\omega > 1$~GeV, and the cut on the transverse momentum of the scattered electron, $p_T > 200$~MeV (cf. Appendix~\ref{sec:app_incl}).

%%%%%%%%%%%%%%%%%%%%%%%%%%%%%%%%%%%%%%%%%%%%%%%%%%%%%%%%%%%%%%%%%%%%%%%%%%%%%%
%%%%%%%%%%%%%%%%%%%%%%%%%%%%%%%%%%%%%%%%%%%%%%%%%%%%%%%%%%%%%%%%%%%%%%%%%%%%%%

\section{Inclusive measurements}
\label{sec:incl}

In the baseline detector configuration, we study the potential for LDMX to make measurements of electron-nucleus processes, the results of which can be used to improve Monte Carlo generators. In this section, we focus on the simplest inclusive measurements LDMX can perform, namely, on the distribution of the scattered electrons on the $(\theta_e, \omega)$ plane, $\theta_e$ and $\omega$ being the scattering angle and the energy transferred to the nucleus, respectively. Until Sec.~\ref{sec:excl}, we do not consider any information on the composition or kinematics of the final-state hadrons.  Here we argue that LDMX will complement the existing knowledge of the inclusive cross sections from the very forward direction to larger scattering angles by providing results for large energy transfers, where they are not available yet~\cite{Benhar:2006er}; see Appendix~\ref{sec:coverage}. 

Our analysis is focused on the fiducial region of the scattered electron's phase space defined by $\omega > 1$ GeV and $p_T > 200$~MeV. This selection is synergistic to the LDMX dark-matter search. Before performing these kinematic selections, we apply parametric angular and momentum/energy smearing of electrons, charged hadrons, and neutral hadrons, according to the expected detector resolutions described above.  We also apply angular acceptance criteria according to the detector acceptance described in Sec.~\ref{sec:LDMX}.  Efficiency effects due to particle identification algorithms are {\it not} applied and require further study.  However, we expect them to be very uniform, well measured, and near unity.

Figure~\ref{fig:elomega} illustrates the distribution of energy transferred by electron to the nucleus for two different selections on its scattering angle $\theta_e$, following the common kinematic selections and energy smearing described above.  In this energy range, all kinematic features in the simulation are broader than the energy resolution. The presented results correspond to the expected number of events for $1\times 10^{14}$~EoT. In the figure, the bands represent statistical uncertainties of the generated Monte Carlo event samples; experimental statistical uncertainties are much smaller than the indicated bands. 

The event distributions obtained using the three generators differ markedly both in the overall rate and in shape.  Exhibiting stronger angular dependence, \gibuu predicts fewer events at large scattering angles than expected according to \genie. 

This behavior is shown in Fig.~\ref{fig:elomega}. In the left panel, corresponding to scattering angles $10^{\circ}\leq\theta_e\leq12.5^{\circ}$, the prediction of \gibuu is smaller by 30\% than that of \genie.  In the right panel, for scattering angles $20^{\circ}\leq\theta_e\leq22.5^{\circ}$, this difference increases to 50\%. While the \genie cross section is dominated by the DIS channel, this is not the case for the \gibuu results, in which resonance excitation is the main mechanism of interaction for energy transfers below 2~GeV, and DIS dominates only at $\omega > 2$~GeV. The largest discrepancies occur at higher-energy transfers ($\omega \gtrsim 2$~GeV, $W^2 \gtrsim 4.4$~GeV$^2$), where events are predominately populated by DIS.  Notably, there are visible differences between the results obtained using different versions of the generators \gibuu and \genie. Nevertheless, they are much less significant than the differences between the predictions of different generators.

In the $20^{\circ}\leq\theta_e\leq22.5^{\circ}$ slice, both the \gibuu and \genie cross sections result entirely from DIS interactions, and agree at a factor of 2 level. \geant, however, deviates significantly from \genie and \gibuu, and the deviation is even larger at higher scattering angles.  This is expected as \geant uses the equivalent-photon approximation to simulate electron-nucleus interactions.  At higher $Q^2$, the exchanged photon becomes highly virtual and this approximation is not valid.  Because of this issue, we do not show \geant predictions in later comparisons.  We note that as \geant is not commonly used as an event generator, the difference between \geant and other generators is not a fair representation of the current modeling uncertainty.  However, the difference between \genie and \gibuu is, and it may even be a conservative estimate on modeling uncertainties.  Comparably large disagreements between \genie and \gibuu are seen in all angular bins, as illustrated in Appendix~\ref{sec:app_incl}.

Notice that, in Fig.~\ref{fig:elomega}, the ranges of electron-scattering angles are narrow and the final energies are well measured. This, combined with precise knowledge of the initial electron energy, makes it possible to accurately control the scattering kinematics, which in turn provides a powerful tool for testing the underlying nuclear and hadronic physics. The large discrepancies between the generator predictions for the double-differential cross section seen in the figure may be less pronounced in more integrated quantities. We explicitly confirmed this by integrating the electron-scattering cross sections for a 4-GeV beam energy over all scattering angles and energy transfers (imposing the same $Q^2>0.03$ GeV$^2$ cut as before). In this case, we find that the predictions of \genie and \gibuu are, in fact, in good agreement. Both generators give 1.9$\times10^{-28} \text{cm}^2$, with the underlying discrepancies completely washed out upon integration.

One has to be mindful about this when interpreting results of neutrino-scattering experiments, where averaging can take place over several variables, including the incoming beam energy. 
As an illustration, consider measurements of pion production induced by charged-current neutrino interactions in the MINERvA experiment, at the kinematics similar to that of DUNE.  The shape of the single differential $d\sigma/dQ^2$ cross sections from Ref.~\cite{McGivern:2016bwh} is reproduced reasonably well by both  \genie~\cite{McGivern:2016bwh} and \gibuu~\cite{Mosel:2017nzk}.

This clearly illustrates a general point: for the purpose of testing the physics models in the generators, detailed measurements of multiply differential cross sections are essential. Fortunately, many such measurements are already available from MiniBooNE~\cite{AguilarArevalo:2010zc}, MINERvA~\cite{Rodrigues:2015hik,Patrick:2018gvi,Gran:2018fxa,Ruterbories:2018gub,Carneiro:2019jds}, MicroBooNE~\cite{Adams:2019iqc}, and T2K~\cite{Abe:2018uhf,Abe:2019sah}, and more can be expected in the future (exploring various semi-inclusive modes). These data, when combined with the electron-scattering measurements described here, will provide a very powerful foundation for generator development. Note that the electron and neutrino measurements are essentially complementary: the electron measurements do not suffer from beam-integration effects, while the neutrino measurements are sensitive to the axial current effects.

LDMX will measure inclusive electron-nucleus scattering rates for energy transfers $1\lesssim\omega\lesssim4$ GeV and scattering angles $5^\circ\lesssim\theta_e\lesssim40^\circ$. With expected $>10^5$ events per bin, the experimental statistical errors will be at subpercent-level.  Instrumental systematic uncertainties are difficult to assess precisely prior to data taking, but the scales of many effects can be estimated by comparison to detailed performance studies of other similar collider and fixed-target experiments. LDMX’s luminosity can be precisely measured by counting incident electrons in the tagging tracker and measuring the target thickness. Electron-reconstruction performance can be quantified precisely using standard candle reactions, such as M{\o}ller scattering. Efficiency uncertainties should be smaller and more uniform than the $\sim$4\% level achieved by the less hermetic CLAS detector (see, e.g., Ref.~\cite{Markov:2019fjy}). Momentum resolution uncertainties should be comparable to the $\sim $3\% achieved at HPS ~\cite{Adrian:2018}, which has a similar detector geometry and beam.  Such resolution would lead to negligible systematic effects on the distributions in Fig.~\ref{fig:elomega}, which vary over much larger energy scales.

Based on these considerations, both statistical and systematic uncertainties are expected to be small, compared with the current theoretical uncertainties.  These features will enable LDMX to discern between \gibuu, \genie, and \geant predictions with high precision, and to perform measurements of the inclusive cross sections for electron scattering on nuclear targets, such as titanium, over a broad kinematics, previously unexplored. Availability of such results is essential for future development and tuning of Monte Carlo generators employed in the long-baseline neutrino-oscillation program.

%%%%%%%%%%%%%%%%%%%%%%%%%%%%%%%%%%%%%%%%%%%%%%%%%%%%%%%%%%%%%%%%%%%%%%%%%%%%%%
%%%%%%%%%%%%%%%%%%%%%%%%%%%%%%%%%%%%%%%%%%%%%%%%%%%%%%%%%%%%%%%%%%%%%%%%%%%%%%

%%%%%%%%%%%%%%%%%%%%%%%%%
\begin{figure}
	\begin{center}
        \includegraphics[width=0.9\columnwidth]{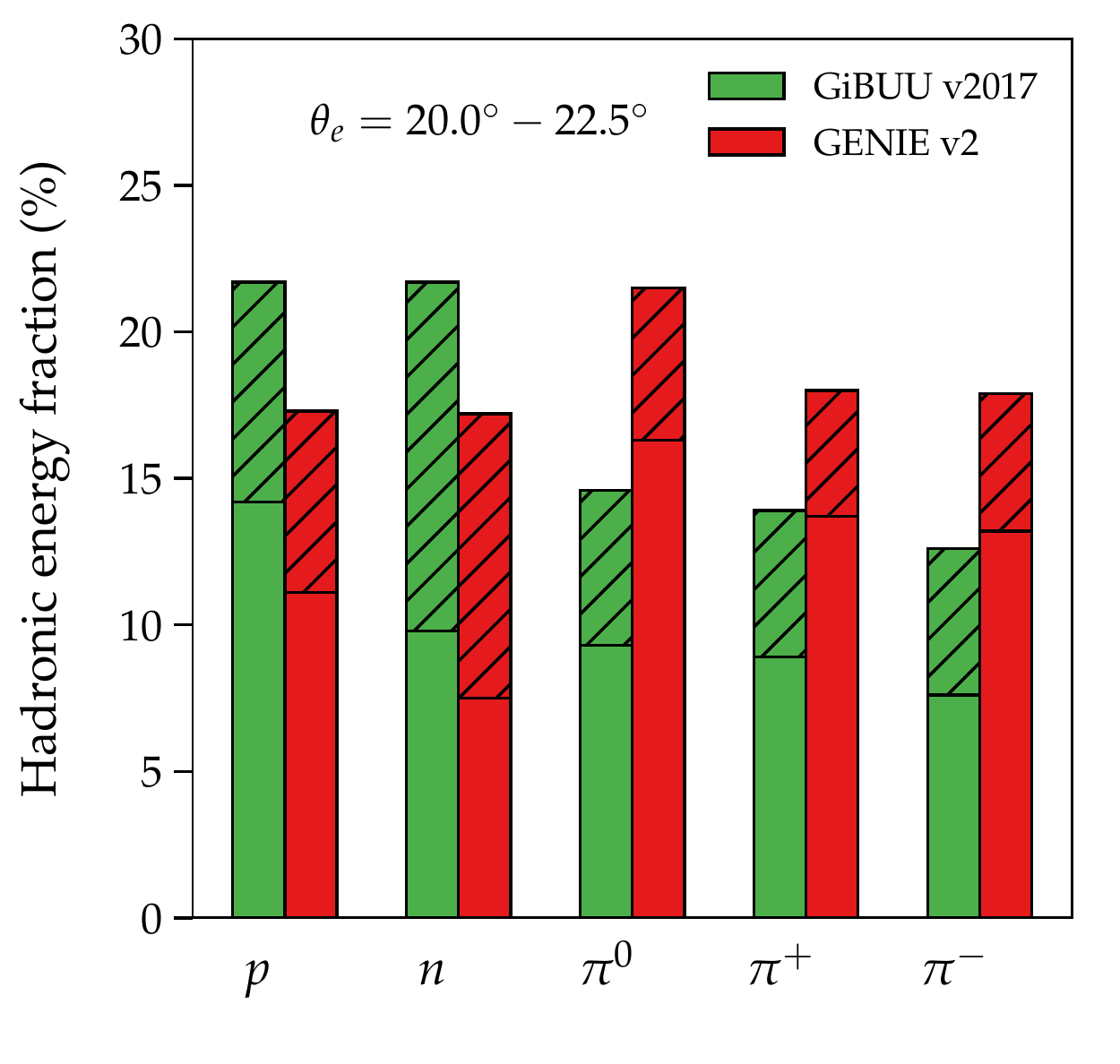}
        \caption{Energy fractions carried by various hadrons in the final state in events with $\omega > 1$~GeV, according to \gibuu and \genie.  The shaded regions illustrate the fractions outside of LDMX acceptance: below detection thresholds or outside the 40$^\circ$ cone. Different compositions of the final hadronic system result from different physics underlying generators.}
        \label{fig:composition}
    \end{center}
\end{figure}
%%%%%%%%%%%%%%%%%%%%%%%%%

\section{(semi)exclusive measurements}
\label{sec:excl}

As explained in Sec.~\ref{sec:e-scattering}, Monte Carlo generators play a~fundamental role in neutrino-energy reconstruction, relating the visible energy---deposited in the detector by the observed particles---with the actual neutrino energy. In order to do so, the contribution of undetected energy---carried away by undetected particles, absorbed in nuclear breakups, etc.~\cite{Friedland:2018vry}---is estimated based on the measured event composition and kinematics. The accuracy of the energy reconstruction relies on the accuracy of the particle multiplicities and spectra predicted by the Monte Carlo simulation. Therefore, availability of precise information on the hadronic final states is essential to validate the models underlying the generators and to estimate their contribution to the systematic uncertainty of energy reconstruction.

Here, we present spectra obtained for coincidence measurements in LDMX of electrons, pions (with particle ID), and neutrons.  We argue that thanks to the angular coverage of LDMX, the measurements can be performed with high efficiency across a broad range of energy and angle.

%%%%%%%%%%%%%%%%%%%%%%%%%
\begin{figure*}
	\begin{center}
        \includegraphics[width=0.95\columnwidth]{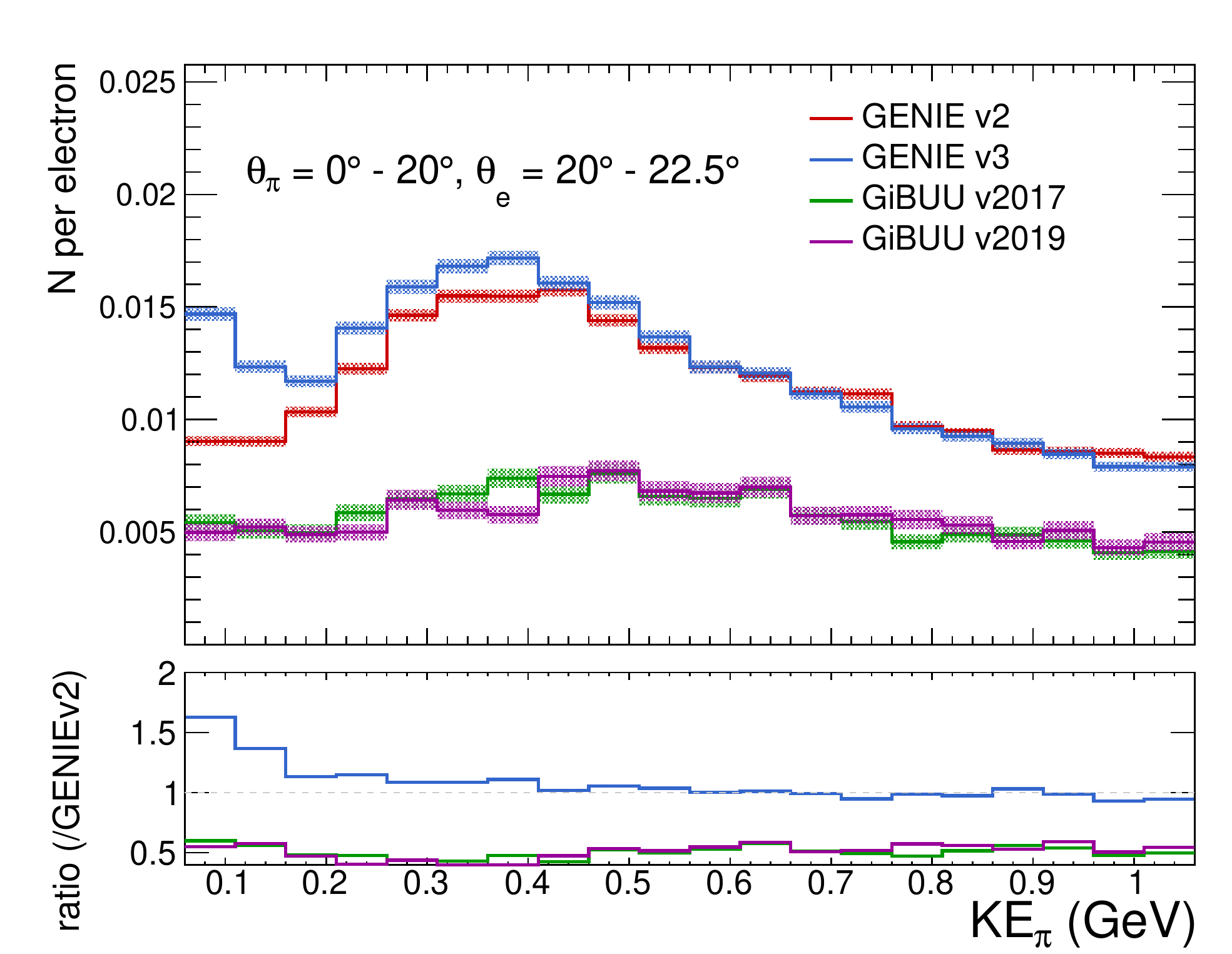}
        \includegraphics[width=0.95\columnwidth]{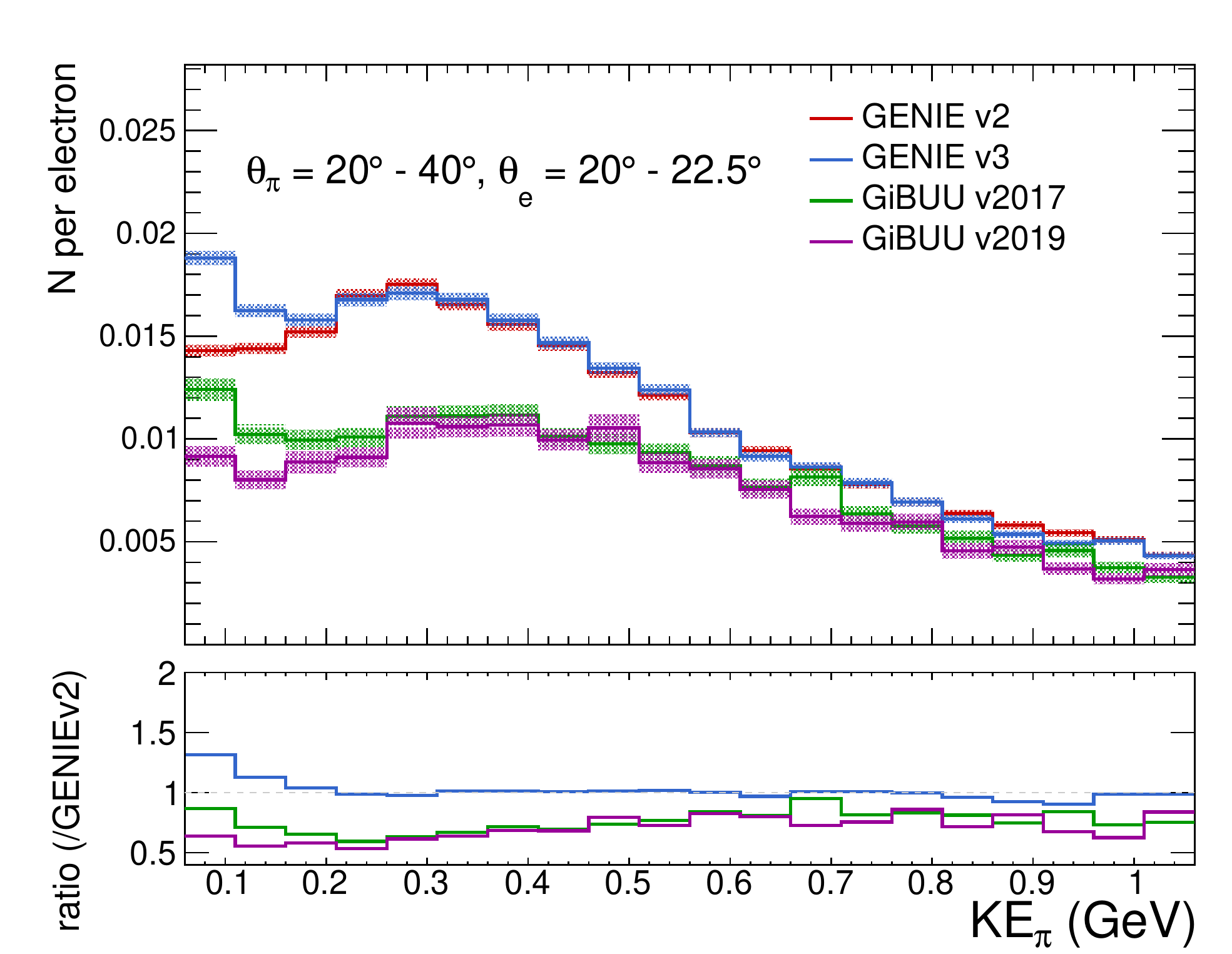}
        \caption{Charged pion kinetic energy distribution after energy/angular resolution smearing with a scattering angle of  $0^\circ\leq\theta_\pi\leq20^\circ$ (left panel) and $20^\circ\leq\theta_\pi\leq40^\circ$ (right panel).  There is an additional selection on the recoiling electron of $\omega > 1$~GeV, $p_T > 200$~MeV, and $20^\circ\leq\theta_e\leq22.5^\circ$.  The pion distributions are presented {\it per electron} within the above electron kinematic selection.  There are approximately $1 \times 10^8$ electrons passing the $\omega$, $p_T$, and $\theta_e$ selections for $1 \times 10^{14}$ electrons on target. } 
        \label{fig:pi}
    \end{center}
\end{figure*}
%%%%%%%%%%%%%%%%%%%%%%%%%

As an example, in Fig.~\ref{fig:composition} we show the energy fraction that goes into different hadronic particles when the electron scattering angle is between 20$^\circ$ and 22.5$^\circ$ and the energy transfer exceeds 1 GeV (corresponding to the right panel of Fig.~\ref{fig:elomega}).  The shaded areas in Fig.~\ref{fig:composition} illustrate the energy fractions that are outside LDMX acceptance, predominantly due to the angular coverage. We observe that most of the final-state particles are within LDMX acceptance.  The neutron acceptance is slightly lower also due to the high threshold, the kinetic energy of 500~MeV.  Even then, LDMX can detect $\sim$50\% neutrons.

The hadronic energy fractions predicted by a generator depend on the interaction channel dominating its total cross section.  While nucleons in the final state carry more energy in the resonance-excitation channel than in DIS, for pions this situation is reversed. Yielding a larger resonance contribution to the total cross section than \genie, \gibuu predicts $\sim$40\% less energy carried by electromagnetic showers initiated by neutral pions, and more energy carried by neutrons. The latter issue is of particular importance because neutrons are particularly difficult to measure in neutrino detectors. If left unresolved, such large discrepancies would result in large uncertainties on the inferred neutrino energy~\cite{Friedland:2018vry}. By measuring these hadronic energy fractions within its geometric acceptance, LDMX will provide a good handle on the relative rate of neutron emission.

\begin{figure}
    \begin{center}
        \includegraphics[width=\columnwidth]{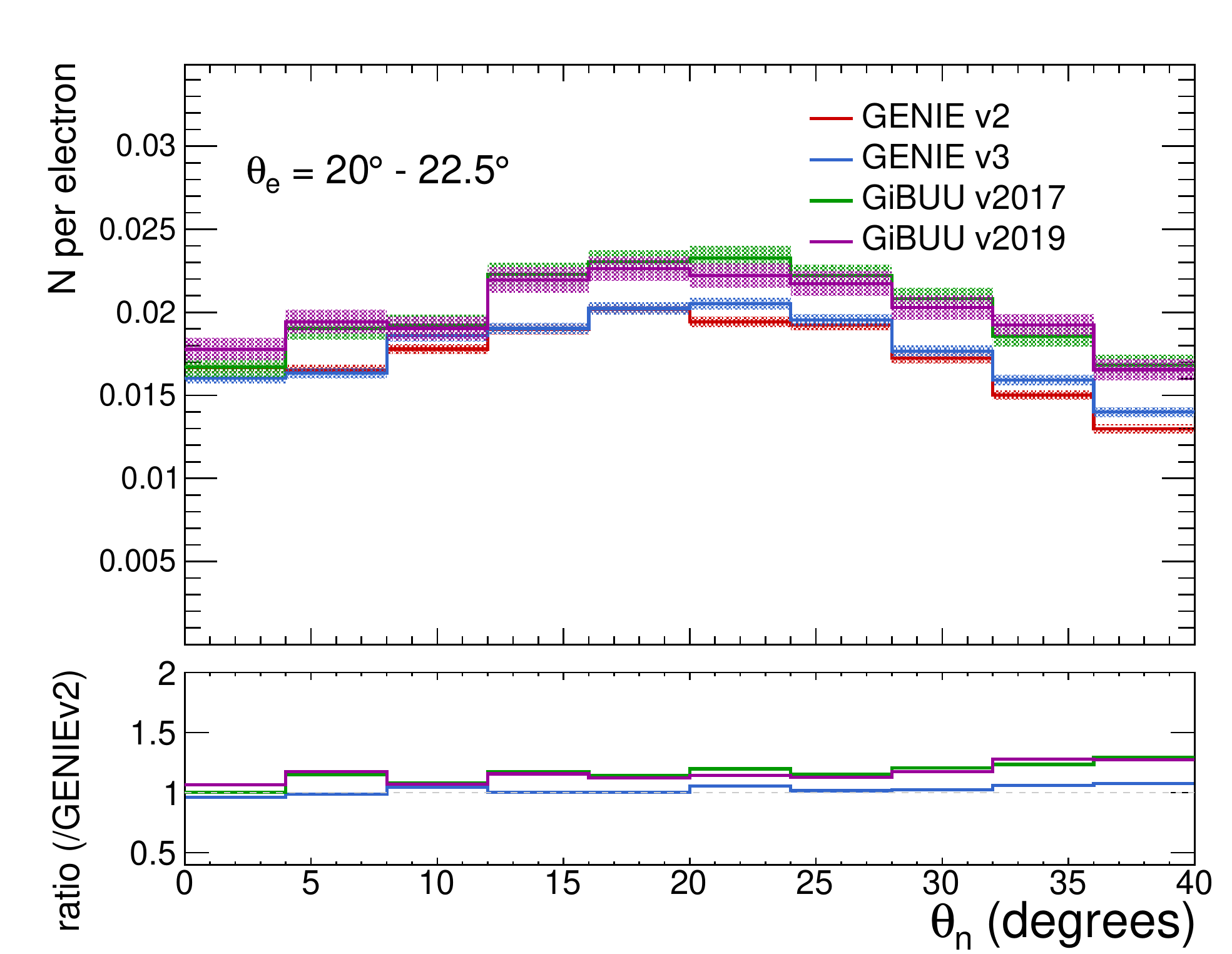}
        \caption{Neutron polar angle distribution after energy/angular resolution smearing.  There is an additional selection on the recoiling electron of $\omega > 1$~GeV, $p_T > 200$~MeV, and $20^\circ\leq\theta_e\leq22.5^\circ$.  The neutron distributions are presented {\it per electron} within the above electron kinematic selection.  There are approximately $1 \times 10^8$ electrons passing the $\omega$, $p_T$, and $\theta_e$ selections for $1 \times 10^{14}$ electrons on target.}
        \label{fig:n}
    \end{center}
\end{figure}

More specifically, the capability of LDMX to measure in coincidence the kinematics of the scattered electron and of the hadronic interaction products is illustrated by the distributions shown in Figs.~\ref{fig:pi} and~\ref{fig:n}. 

%%%%%%%%%%%%%%%%%%%%%%%%%
%%%%%%%%%%%%%%%%%%%%%%%%%

Figure~\ref{fig:pi} presents the pion kinetic energy distributions expected in LDMX when the corresponding electron kinematics is selected in a similar manner as in the previous section: $\omega > 1~$GeV, $p_T > 200$~MeV, and $20^\circ\leq\theta_e\leq22.5^\circ$. We expect approximately $1 \times 10^8$ electrons with that particular kinematic selection for $1 \times 10^{14}$ electrons incident on the target. After accounting for the acceptance and energy resolution of the tracker, LDMX can measure the charged-pion kinetic energy down to $\sim$60~MeV.  We present the distribution up to 1~GeV, where LDMX is expected to have good pion/proton discrimination.

The distributions in Fig.~\ref{fig:pi} are normalized per electron meeting the selection criteria, in order to remove the generator differences for inclusive electron scattering discussed in Sec.~\ref{sec:incl}.  We see that \genie predicts more pions, about a factor of 2 more in the forward region, while \gibuu yields a slightly harder pion spectrum. 

Similarly to the electron case, the pion energy resolution is sufficiently small that its effect is invisible in the figure, and features in pion spectra predicted by generators, e.g., the peak toward lowest pion energies due to final-state interactions,  are preserved.  We also observe a sensitivity to the difference between the pion spectra for $0^\circ\leq\theta_\pi \leq 20^\circ$ and for $20^\circ\leq\theta_\pi\leq40^\circ$, illustrating the advantage of having fine-grained tracking detector for all charged particles.

In Fig.~\ref{fig:n}, the angular distributions of all neutrons in an event within the acceptance of the tracker and calorimeter and with (smeared) kinetic energies greater than 500~MeV are shown.  Again, this is with the same selection on the electron as in the pion result.  The distributions show large overall rate differences between the generators, but even within the shape of the distributions, there are differences at the 30\%--40\% level.

From the representative distributions we have shown for the electron and hadron kinematics, it is clear that there are large deviations in the predictions of electron-nucleus interactions from various state-of-the-art generators.  LDMX will provide good measurements of these multiparticle final states.  Figures~\ref{fig:pi} and~\ref{fig:n} show the pion kinetic energies and neutron angular distributions {\it per incoming electron within a narrow angular slice}, but as is noted above, we expect approximately $1 \times 10^8$ electrons with that kinematic selection.  Therefore, the per-bin statistical uncertainties on these measurements will be at the percent level or smaller.

The systematic uncertainties discussed in the context of inclusive measurements translate directly to the case of (semi)exclusive measurements.  The main new systematic in this case is the efficiency and cross-contamination of hadron particle identification using $dE/dx$. For 1~GeV and below, the rate of cross-contamination for charged pions and protons is likely to be similar to the several percent level observed at CMS~\cite{Khachatryan:2010pw, Giammanco:2011zz}; this sets the scale for a conservative estimate of the systematic uncertainties as well.  Contamination from kaons, due to their much lower absolute rate, is expected to be even less than from protons and pions.  For neutron identification, the detector technology chosen, scintillator-based sampling calorimetry, is quite mature. While the readout technology and geometry is different, the CMS experiment measures neutral hadrons down to the GeV scale and the uncertainties on energy measurements are at the $\sim$10\%--20\% level~\cite{Chatrchyan:2009ag}.

To summarize, similarly to the inclusive case, the expected statistical and systematic errors will be sufficiently small to enable precise measurements of (semi){\-}exclusive electron-nucleus cross sections, by detecting final-state hadrons in coincidence with scattered electron. This data will be vital to understanding neutrino-nucleus interactions and event reconstruction at DUNE.  Furthermore, it is important to note that there is very little existing data for exclusive measurements of neutron knockout induced by electron-nucleus scattering, and thus, any such measurements will be important to constrain Monte Carlo models.

%%%%%%%%%%%%%%%%%%%%%%%%%%%%%%%%%%%%%%%%%%%%%%%%%%%%%%%%%%%%%%%%%%%%%%%%%%%%%%
%%%%%%%%%%%%%%%%%%%%%%%%%%%%%%%%%%%%%%%%%%%%%%%%%%%%%%%%%%%%%%%%%%%%%%%%%%%%%%

\section{Future potential}
\label{sec:future}

In the baseline dark-matter configuration and nominal running, LDMX can be expected to perform valuable measurements of both inclusive and (semi)exclusive electron scattering on nuclear targets of interest for DUNE. Here we enumerate potential ways, some more challenging to realize than others, to extend the physics program beyond the nominal one:
%%%%%%%%%%%%%%%%%%%%%%%%%
\begin{itemize}
	\item The nominal physics selections can be extended to smaller energy transfers to fully cover the regions in which resonance-production and meson-exchange currents provide important contributions to the cross section.  However, there are challenges with triggering on this topology (prescaling is a possibility) and eventually also issues of detector resolution.  More study is left for future work to understand the impact of such measurements.
	\item In this analysis we assume a 4-GeV electron beam, but there is potential for extending measurements to higher energies. In particular, an 8-GeV electron beam from LCLS-II will move the LDMX acceptance contours to the right in Fig.~\ref{fig:DUNE_distribution}.  This would allow to cover more of the DIS phase space with relatively little change in the detector configuration.
	\item Varying the target material would provide more data for nuclear modeling, allowing for deeper understanding of the cross-sections' dependence on the atomic number. While a dedicated study is necessary to make a conclusive statement, it may be possible to employ an argon target, which would directly address the needs of the neutrino community. Measurements for helium, deuterium, and hydrogen are also of great importance, as they would provide a handle on the effect of nuclear transparency on the exclusive cross sections and cleanly separate hadronic and nuclear effects. A scintillator target could also be considered. As in these cases there is some potential conflict with the dark-matter program, they may require dedicated beam time.
	\item In order to improve energy acceptance for low-energy charged particles, the dipole magnetic field can be reduced.  The effect of a reduced magnetic field on the reconstruction of higher-energy particles is left to study in future work.
	\item Although all the generator differences discussed here are manifest (at least in part) in the forward region, it would be ideal to simultaneously constrain the hadronic energy covering also wider angles.  This could be achieved by the combination of LDMX data and e4nu CLAS data.  It is also possible to install a wide-angle detector in front of LDMX, to record both types of information at an event-by-event level.
	\item Additional detector systems such as improved silicon tracking or high-angle scintillating detectors could improve the angular acceptance of LDMX for electron-nucleus measurements.  Their benefits and potential costs, including the effect on the dark-matter program, will require further study.
\end{itemize}
%%%%%%%%%%%%%%%%%%%%%%%%%

%%%%%%%%%%%%%%%%%%%%%%%%%%%%%%%%%%%%%%%%%%%%%%%%%%%%%%%%%%%%%%%%%%%%%%%%%%%%%%
%%%%%%%%%%%%%%%%%%%%%%%%%%%%%%%%%%%%%%%%%%%%%%%%%%%%%%%%%%%%%%%%%%%%%%%%%%%%%%

\section{Summary and Conclusions}

Modern neutrino experiments depend on the ability of event-generator codes to accurately model scattering of neutrinos of several-GeV energies on nuclear targets. This includes predicting both inclusive cross sections and the properties of the final-state hadronic system. This is a very challenging problem, as both nonperturbative hadronic and nuclear effects operate in this energy range and must be simultaneously accounted for. No \emph{ab initio} treatment encompassing all this physics is presently available. The task of building a reliable event generator is thus an art as much as a science, combining a number of models in ways that fairly reflect the underlying physics and pass a battery of experimental tests.

Given this state of affairs, direct data comparisons are absolutely essential for validating and improving today’s generators. In such comparisons, electron-scattering experiments have a very important role to play. They complement what might be learned from neutrino detectors in several important ways, among which are high event rates and precisely known kinematics. This point has been recognized in the neutrino community~\cite{Gallagher:2004nq} and modern event generators are built to model neutrino-nucleus and electron-nucleus interactions using common physics frameworks.

There exists another reason why electron-scattering experiments are of interest to modern particle physics: they offer a laboratory for testing theoretical ideas about dark sectors. The LDMX experiment, in particular, has been conceived for just such a purpose and its design has been optimized for searching for sub-GeV dark matter with unprecedented reach. It turns out, as we argue in this paper, that the two seemingly unrelated tasks are in reality highly synergistic and LDMX will provide invaluable data on electron-nucleus scattering processes that can be very helpful for the neutrino-oscillation program. With a 4-GeV electron beam, LDMX would be able to probe a region of DUNE's scattering phase space where the event density is high (cf. Fig.~\ref{fig:channels} in Appendix~\ref{sec:DUNE_distri}), the theoretical description is challenging, and the existing data coverage is very limited (cf. Fig.~\ref{fig:coverage} in Appendix~\ref{sec:coverage}).

To quantify this statement, we compared predictions of \genie and \gibuu, two of the leading event generators on the market. We argued that both statistical and systematic errors achievable at LDMX are expected to be significantly smaller than the differences between the predictions of these generators (cf. Fig.~\ref{fig:elomega-suppl} in Appendix~\ref{sec:app_incl}). This applies not only to inclusive cross sections, but also to measurements of specific hadronic final states. In fact, LDMX will be able to perform high-resolution studies of spectra and angular distributions for a variety of interaction products---making use of its capability of measuring electrons, photons, neutrons, pions, and protons---over a large geometrical acceptance with high efficiency. These measurements will improve our understanding of hadronic physics in the theoretically challenging region of transition from resonance excitations to deep-inelastic scattering. Moreover, LDMX has also good acceptance and resolution of neutrons, which are a crucial source of missing energy in neutrino detectors. LDMX can thus serve as an important tool in constraining the neutrino-nucleus cross-section uncertainties that plague the neutrino-oscillation program.

For all of these reasons, we strongly encourage the LDMX Collaboration to pursue detailed modeling studies of the scattering processes outlined in this paper and to include the corresponding measurements in future data taking.

%%%%%%%%%%%%%%%%%%%%%%%%%%%%%%%%%%%%%%%%%%%%%%%%%%%%%%%%%%%%%%%%%%%%%%%%%%%%%%
%%%%%%%%%%%%%%%%%%%%%%%%%%%%%%%%%%%%%%%%%%%%%%%%%%%%%%%%%%%%%%%%%%%%%%%%%%%%%%

\begin{acknowledgments}
We are grateful to the LDMX Collaboration for useful discussions and comments.  We also thank the developers of the {\sc ldmx-sw} code for simulations used in this study.  We thank Steven Dytman for assistance with using \genie version 2 in electron-scattering mode, and Steven Gardiner and Afroditi Papadopoulou for help with \genie version 3 and for providing the $e$-Ti cross section spline. We express our gratitude to Ulrich Mosel for help in running \gibuu.  A.~M.~A, A.~F., O.~M., P.~S., N.~Toro, and S.~W.~L. are supported by the U.S. Department of Energy under Contract No. DE-AC02-76SF00515.  N.~Tran is supported by Fermi Research Alliance, LLC under Contract No. DE-AC02-07CH11359 with the U.S. Department of Energy, Office of Science, Office of High Energy Physics.
\end{acknowledgments}

%%%%%%%%%%%%%%%%%%%%%%%%%%%%%%%%%%%%%%%%%%%%%%%%%%%%%%%%%%%%%%%%%%%%%%%%%%%%%%
%%%%%%%%%%%%%%%%%%%%%%%%%%%%%%%%%%%%%%%%%%%%%%%%%%%%%%%%%%%%%%%%%%%%%%%%%%%%%%

\appendix

%%%%%%%%%%%%%%%%%%%%%%%%%
\begin{figure}
    \begin{center}
        \includegraphics[width=\columnwidth]{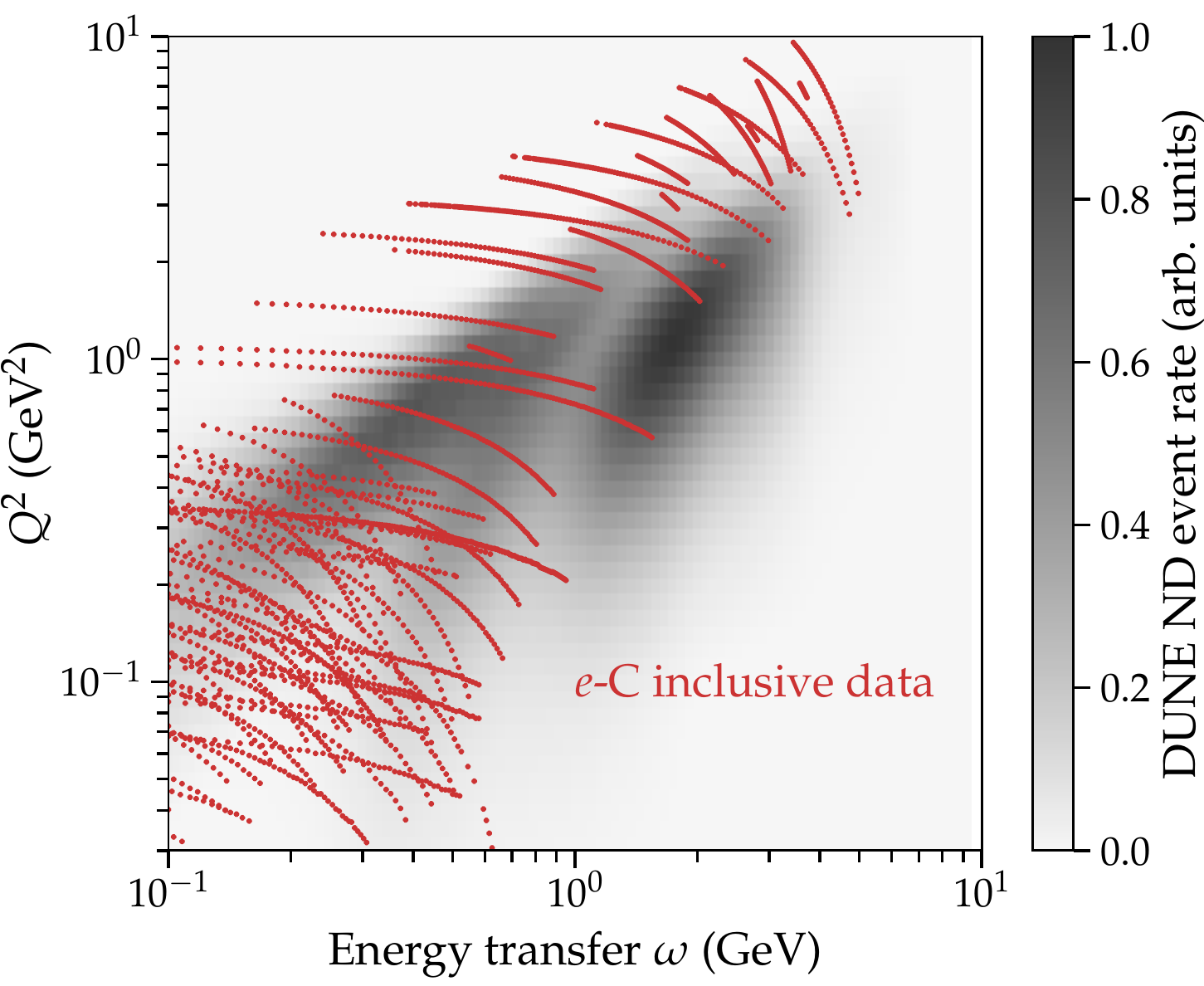}
        \caption{Existing data for inclusive electron scattering on carbon~\cite{Whitney:1974hr,Barreau:1983ht,O'Connell:1987ag,Bagdasaryan:1988hp,Baran:1988tw,Sealock:1989nx,Day:1993md,Arrington:1995hs,Arrington:1998hz,Arrington:1998ps,Fomin:2008iq,Fomin:2010ei,Dai:2018xhi}, overlaid on the simulated distribution of charged-current $\nu_\mu$ events in the DUNE near detector.}
        \label{fig:coverage}
    \end{center}
\end{figure}
%%%%%%%%%%%%%%%%%%%%%%%%%

%%%%%%%%%%%%%%%%%%%%%%%%%
\begin{figure*}
    \begin{center}
        \includegraphics[width=0.49\textwidth]{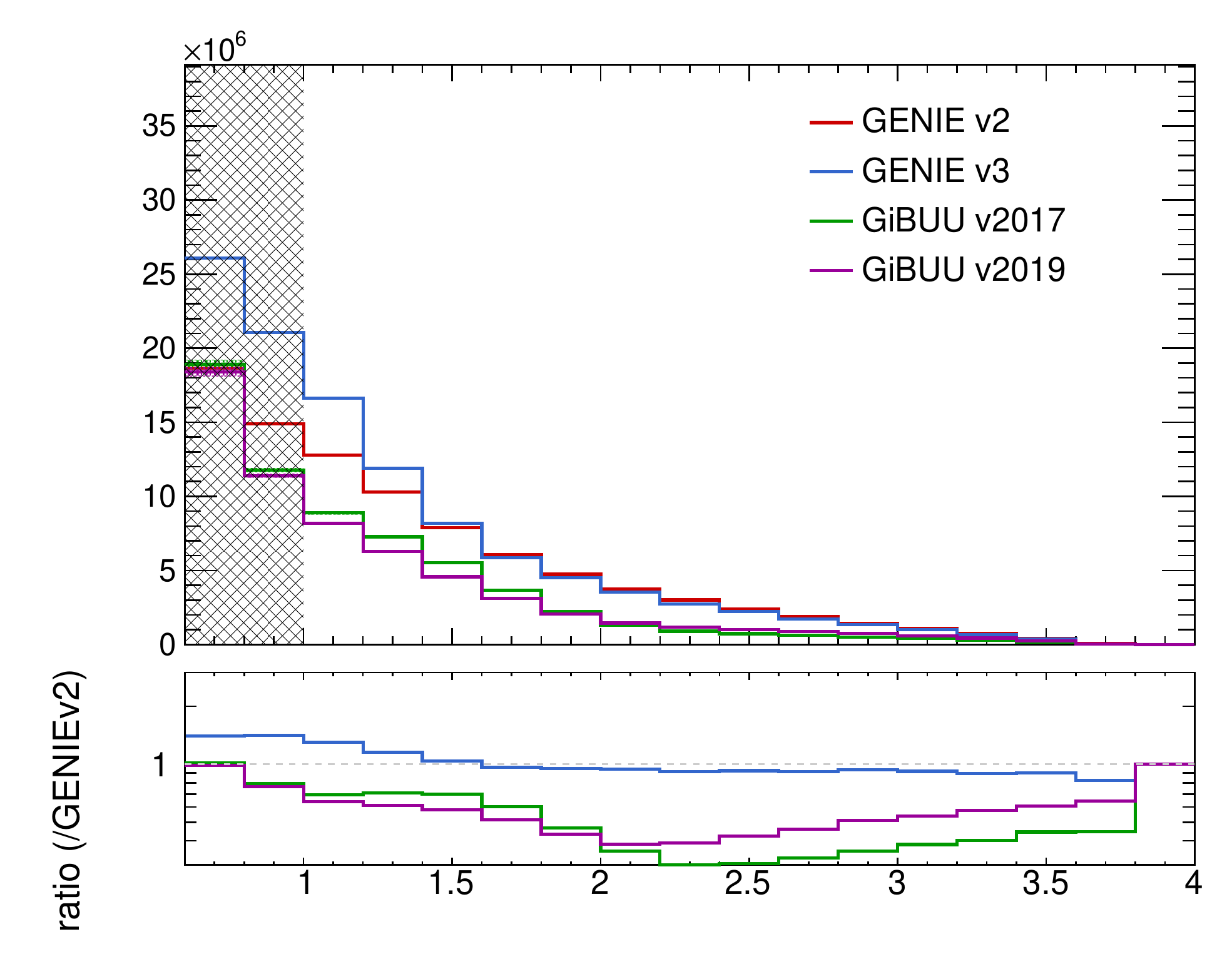}
        \includegraphics[width=0.49\textwidth]{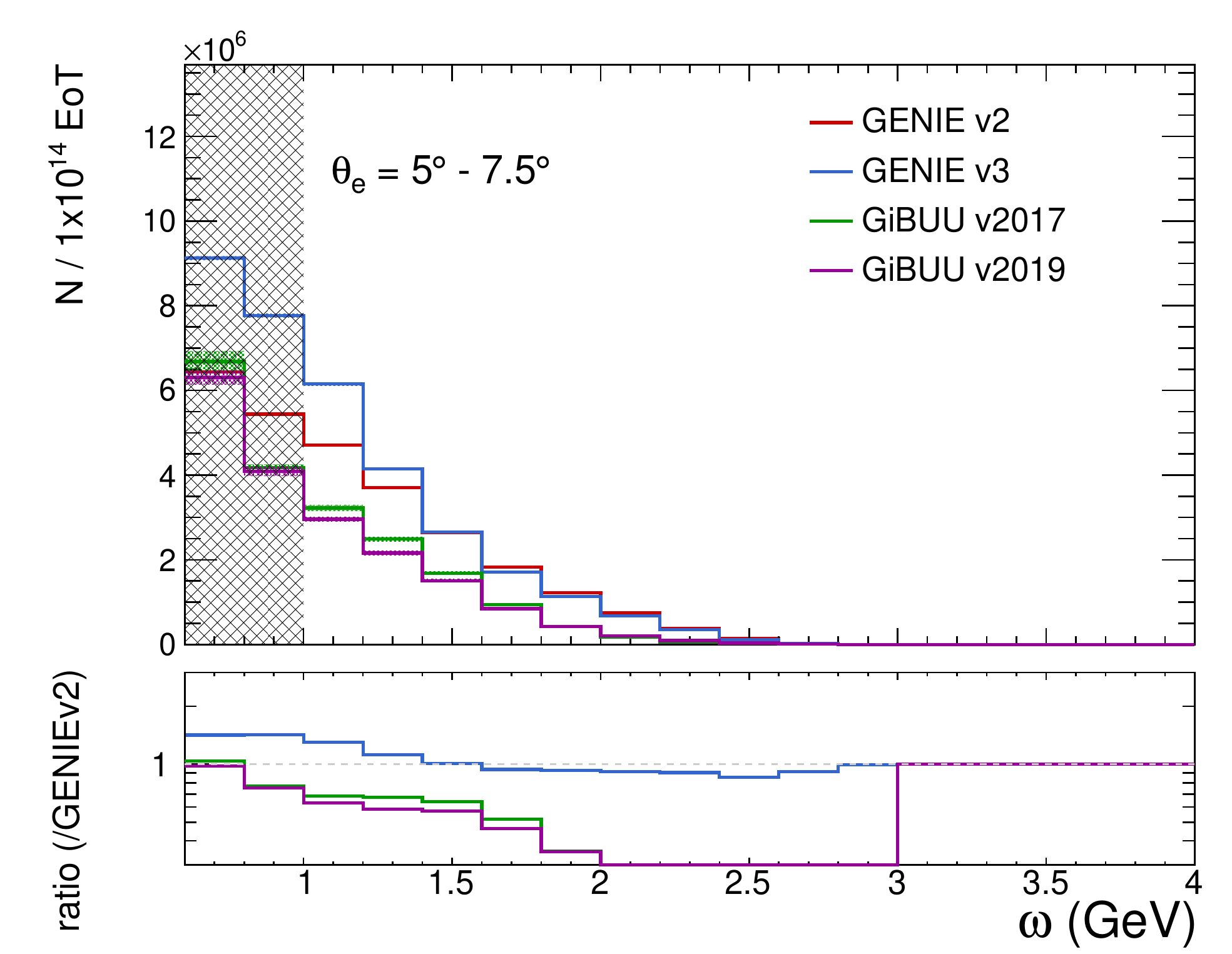} \\
        \includegraphics[width=0.49\textwidth]{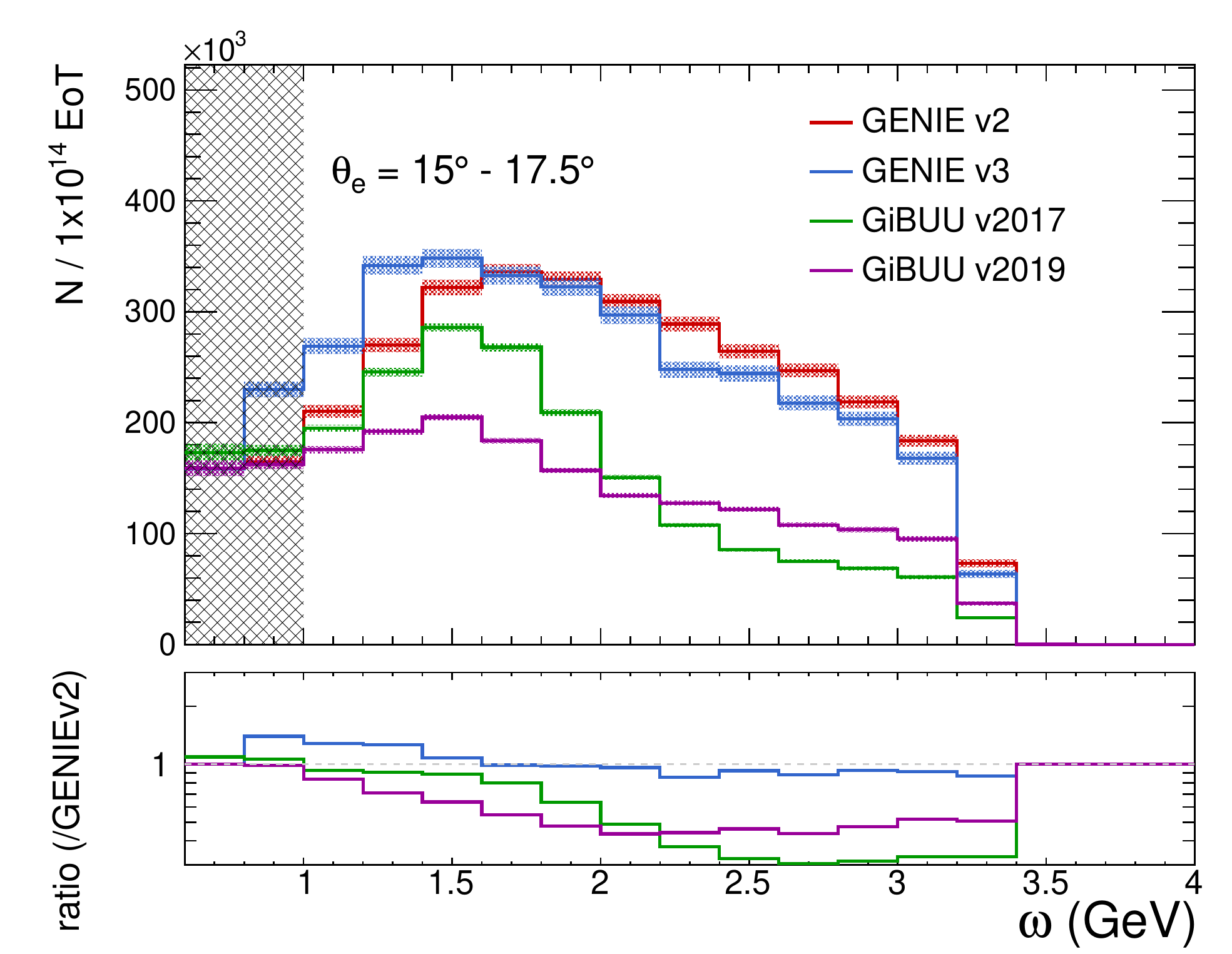}
        \includegraphics[width=0.49\textwidth]{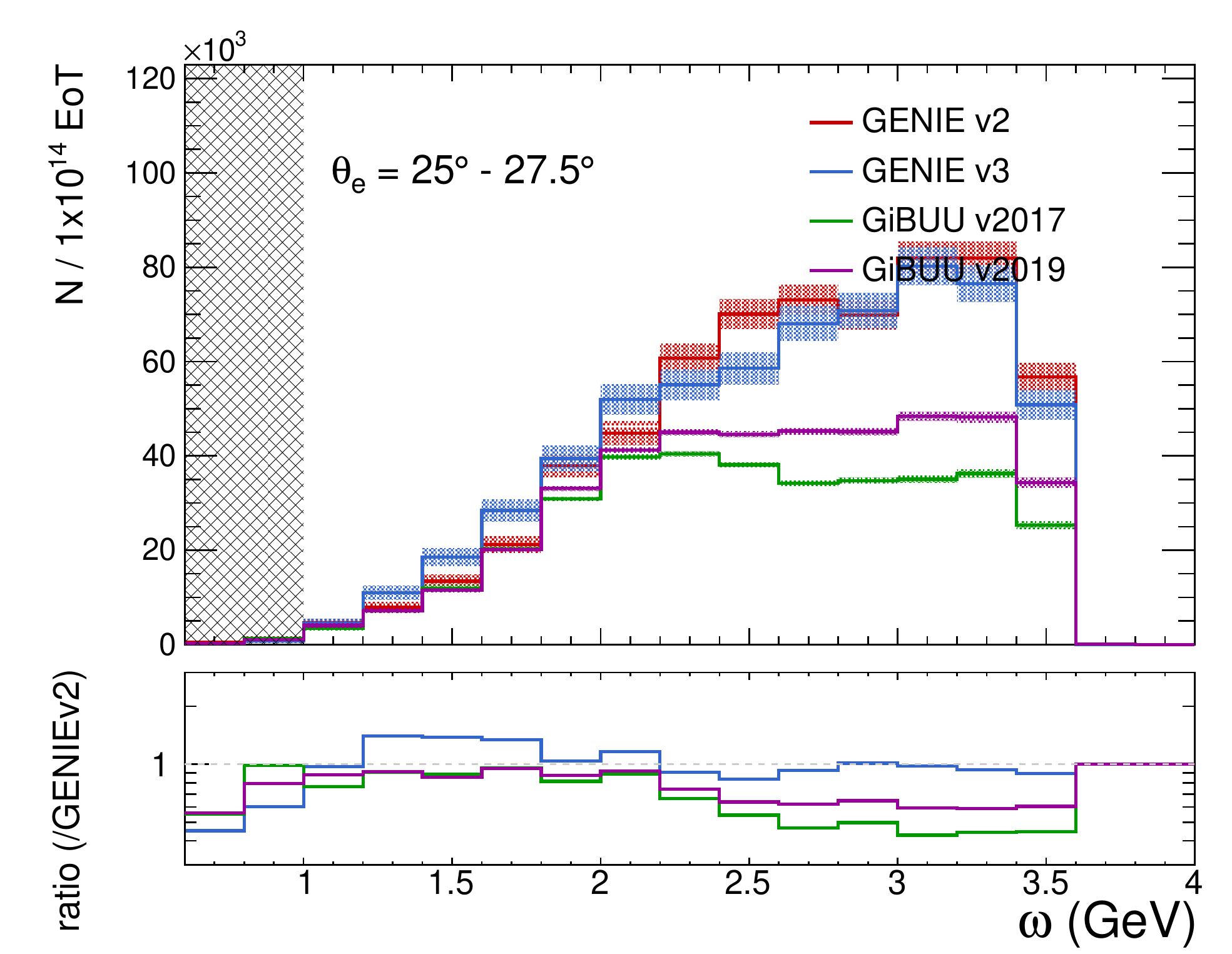} \\
        \includegraphics[width=0.49\textwidth]{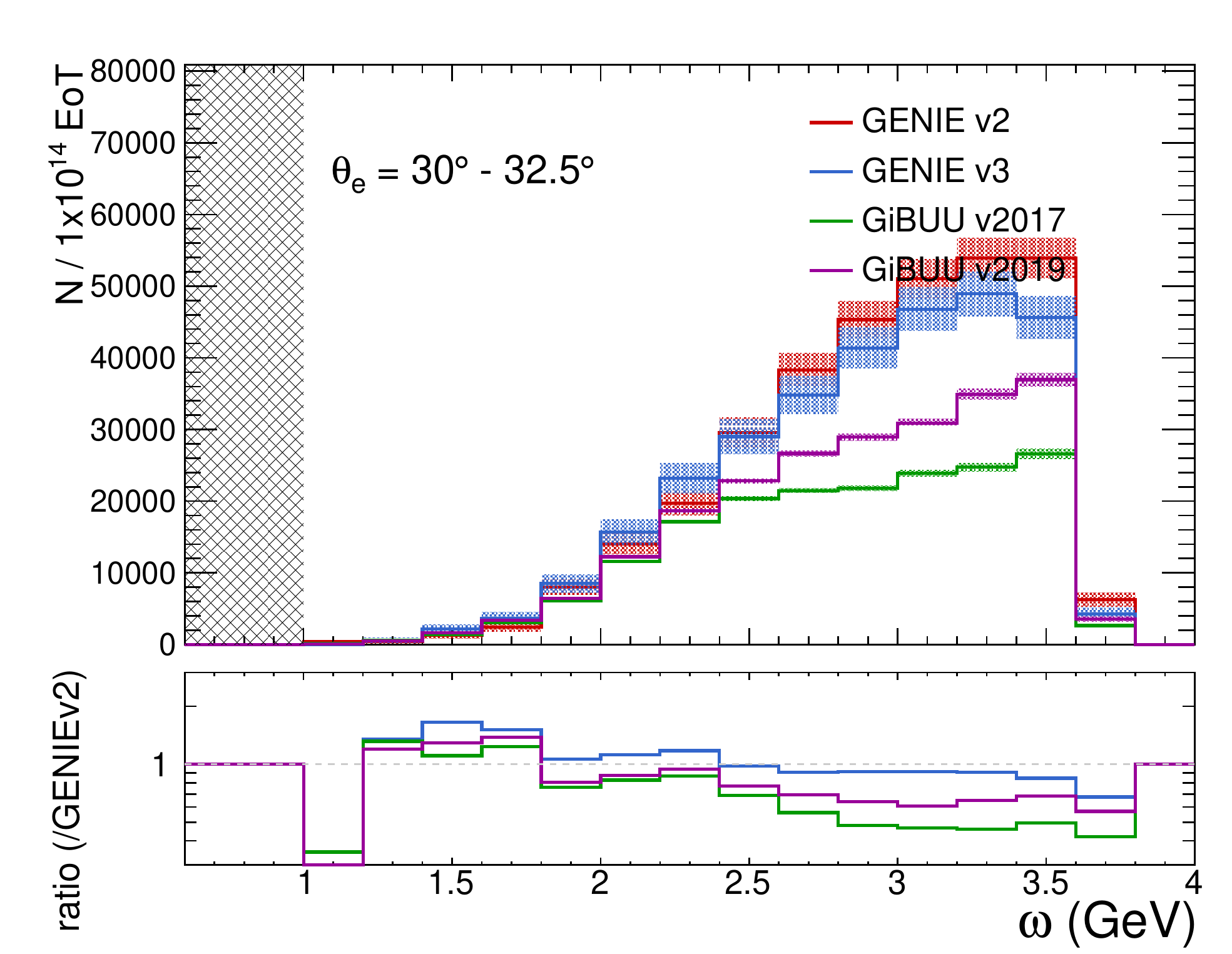}
        \includegraphics[width=0.49\textwidth]{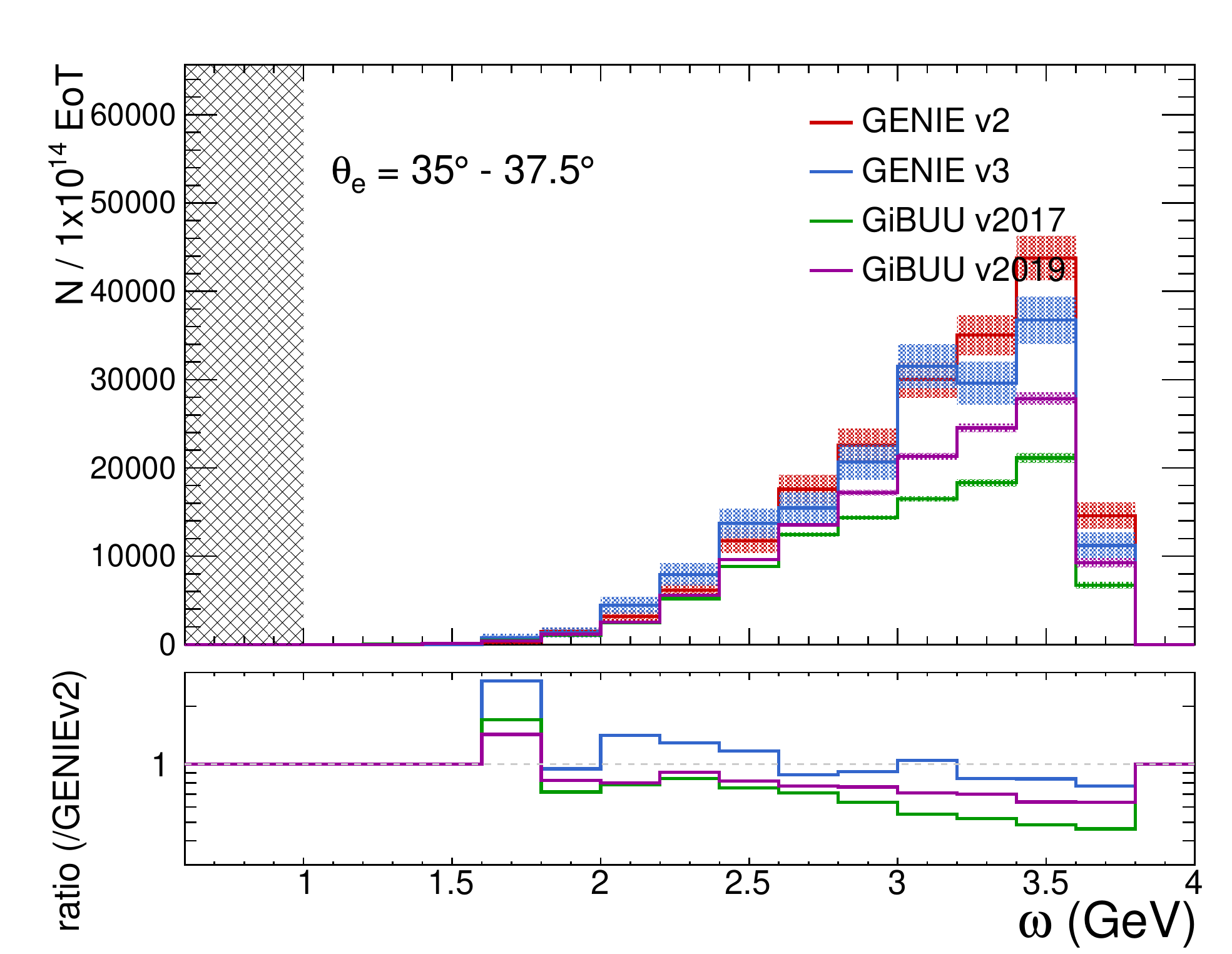}
        \caption{Event distribution as a function of electron energy transfer, $\omega$, for the scattering angles in the range 0$^{\circ}$--40$^{\circ}$ (top left), 5.0$^{\circ}$--7.5$^{\circ}$ (top right),
        15.0$^{\circ}$--17.5$^{\circ}$ (middle left),
        25.0$^{\circ}$--27.5$^{\circ}$ (middle right),
        30.0$^{\circ}$--32.5$^{\circ}$ (bottom left), and
        35.0$^{\circ}$--37.5$^{\circ}$ (bottom right). Note that electrons are required to have the transverse momentum exceeding 200~MeV.}
        \label{fig:elomega-suppl}
    \end{center}
\end{figure*}
%%%%%%%%%%%%%%%%%%%%%%%%%

%%%%%%%%%%%%%%%%%%%%%%%%%
\begin{figure*}
    \begin{minipage}{0.48\textwidth}
        \includegraphics[width=0.9\textwidth]{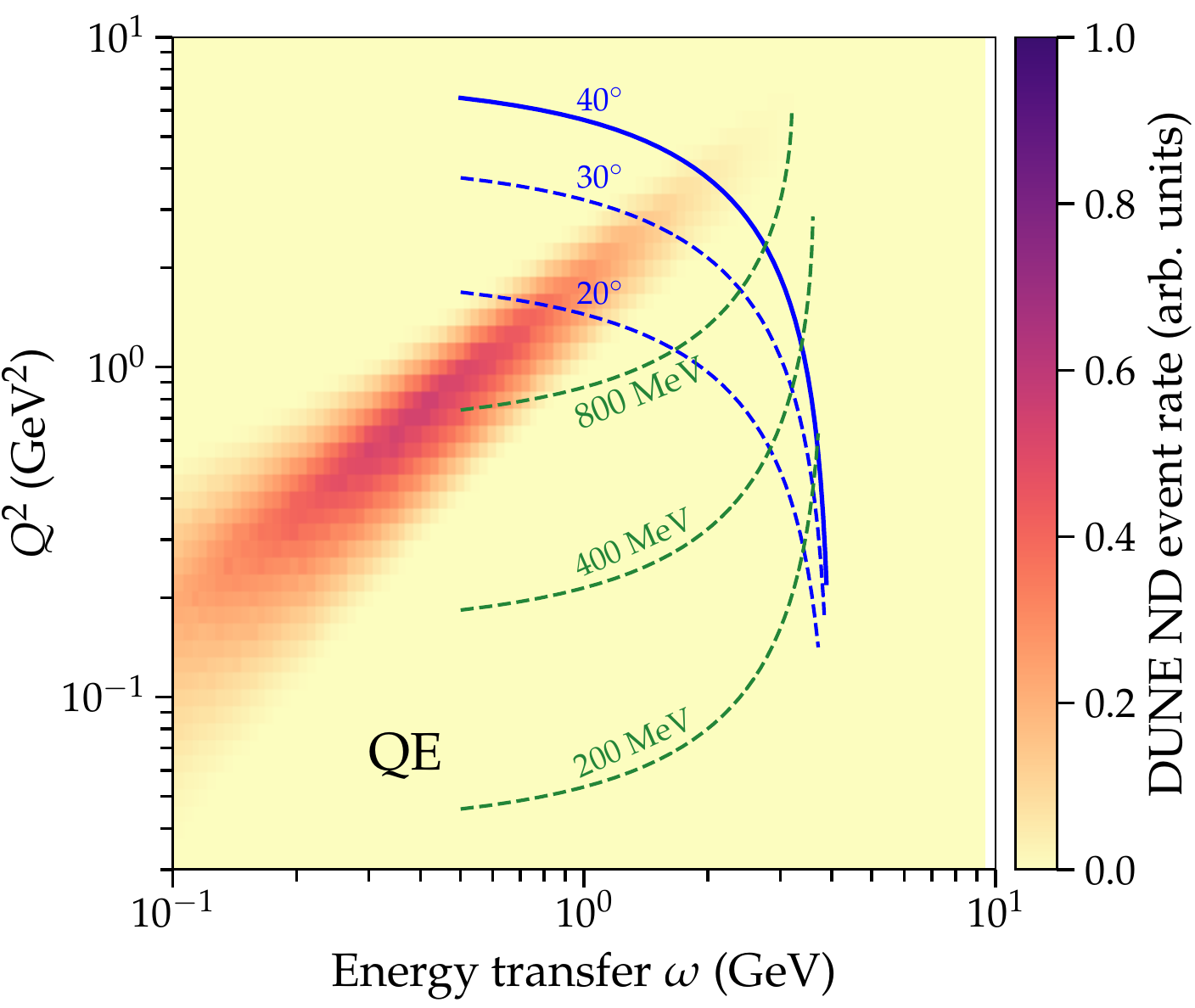}
    \end{minipage}
    \begin{minipage}{0.48\textwidth}
        \includegraphics[width=0.9\textwidth]{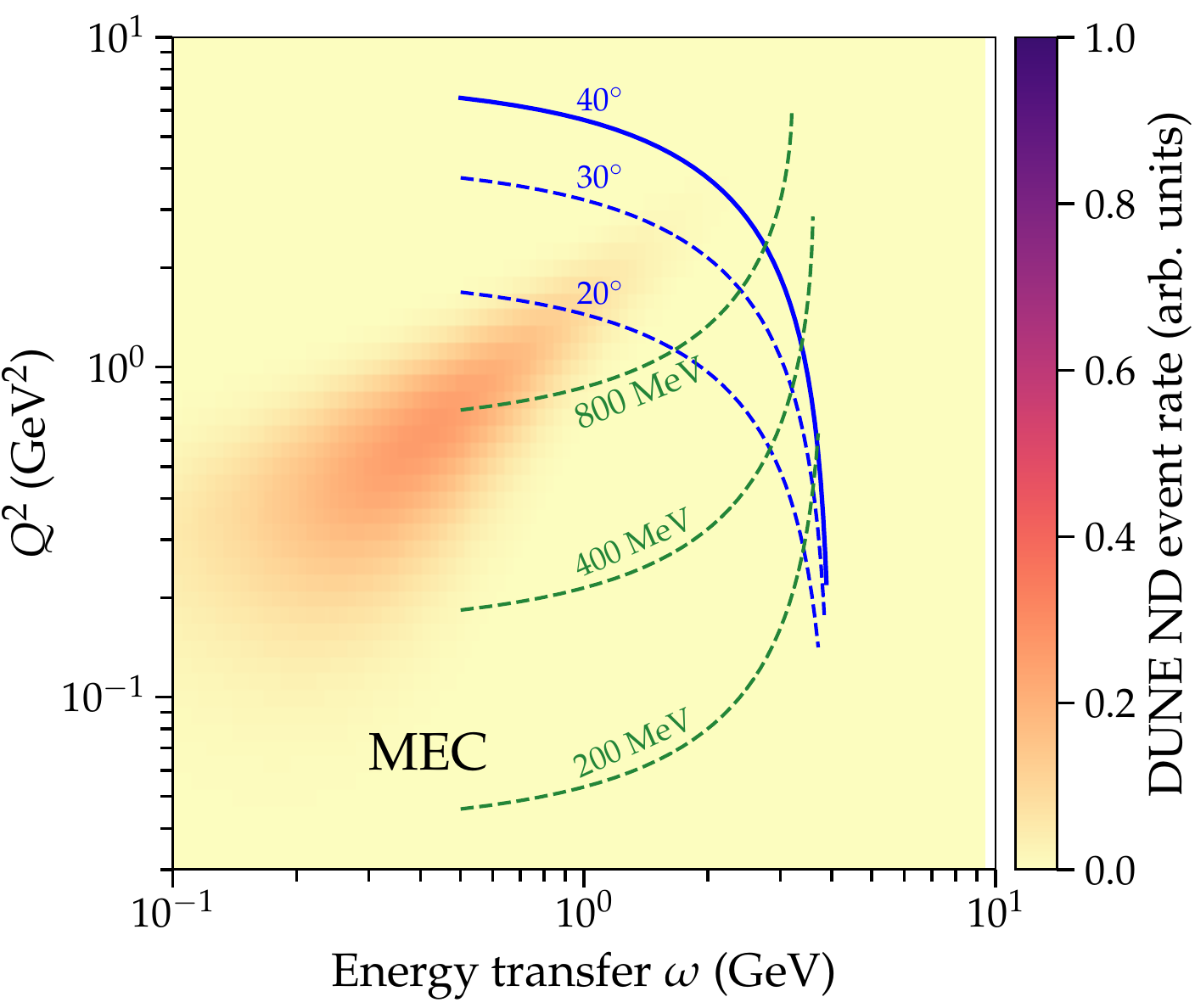}
    \end{minipage}
    \\\vspace{1em}
    \begin{minipage}{0.48\textwidth}
        \includegraphics[width=0.9\textwidth]{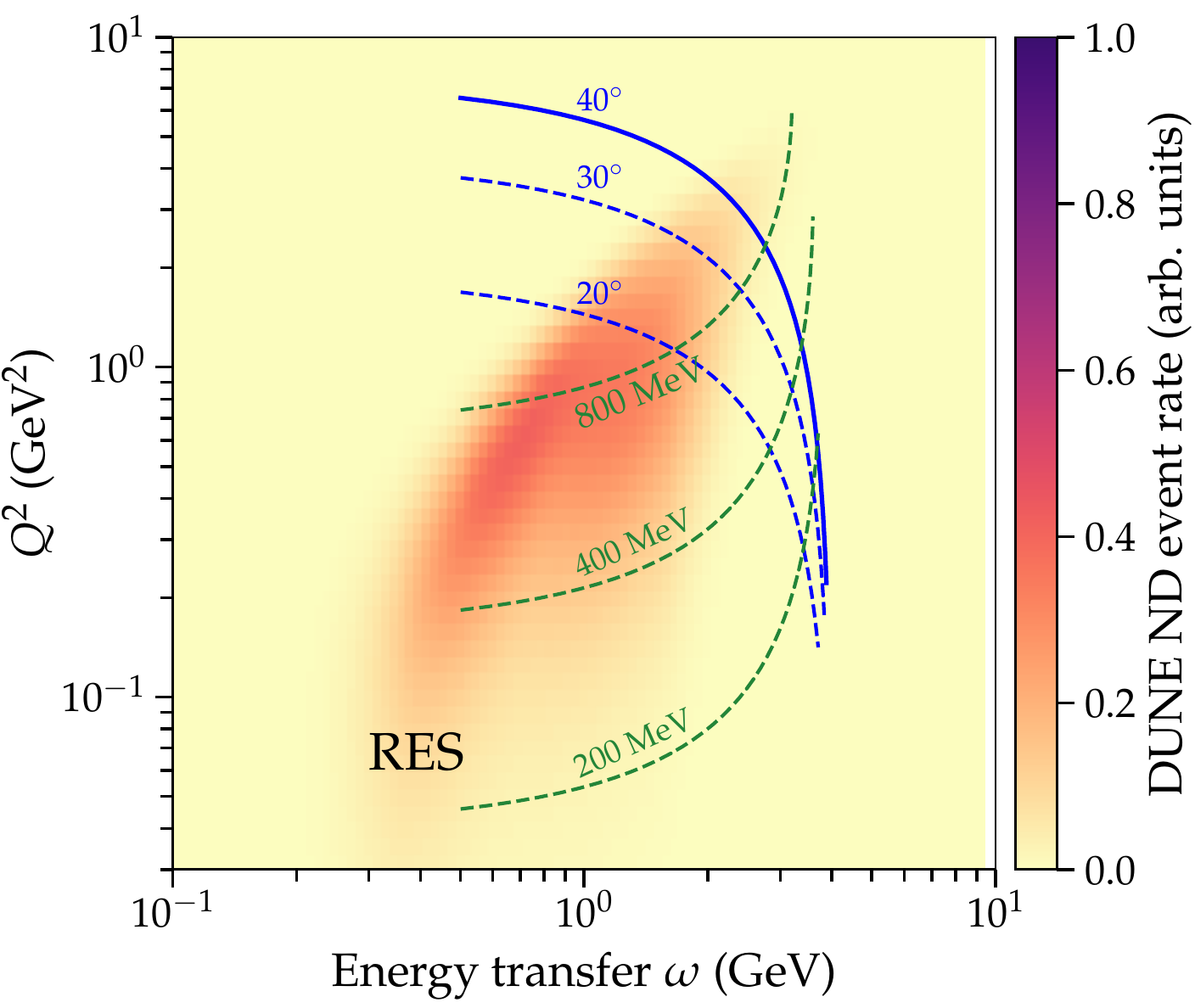}
    \end{minipage}
    \begin{minipage}{0.48\textwidth}
        \includegraphics[width=0.9\textwidth]{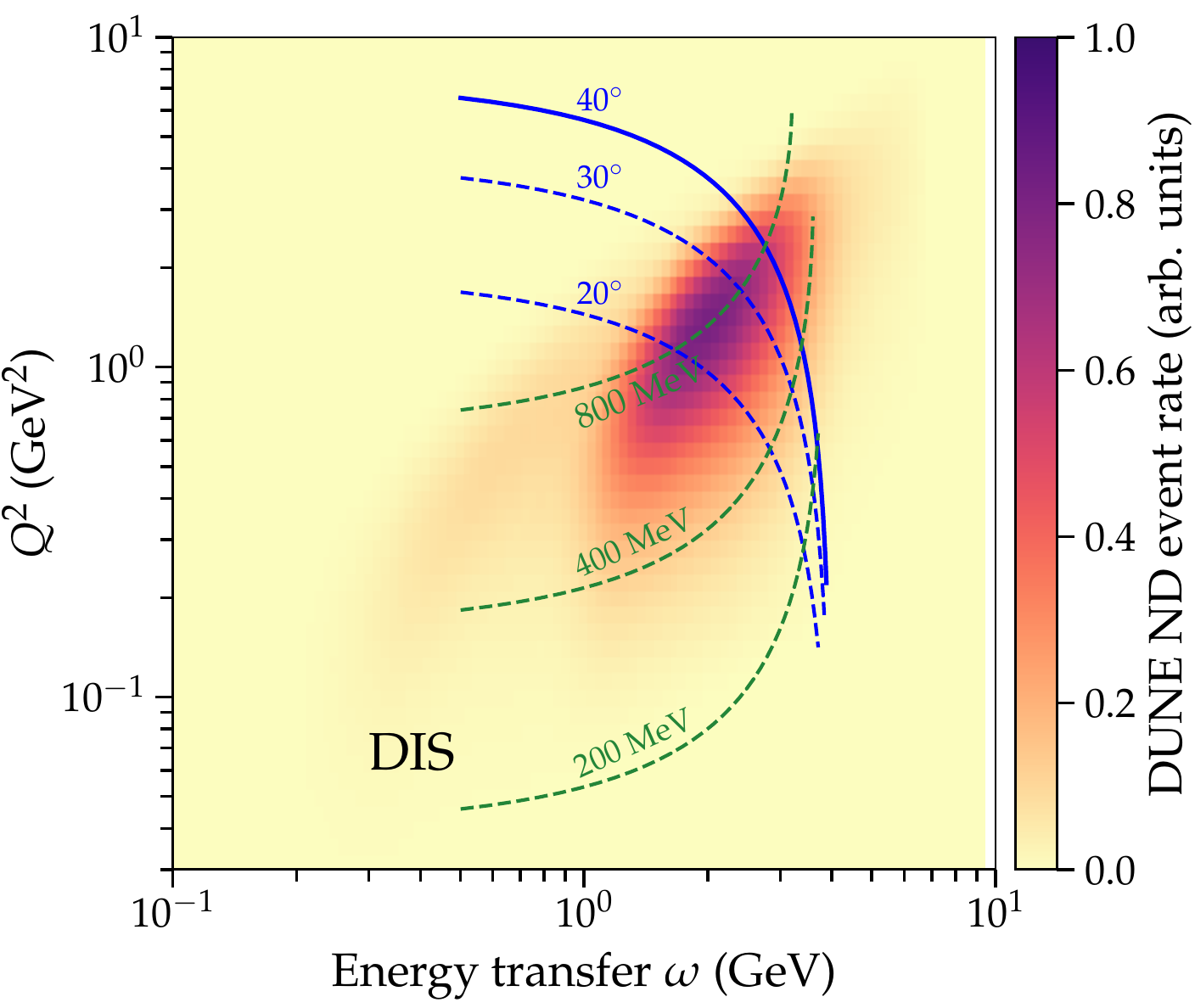}
    \end{minipage}
        \caption{Event distributions in the DUNE near detector according to \gibuu, broken into individual interaction channels: QE, MEC, RES, and DIS.}
        \label{fig:channels}
\end{figure*}
%%%%%%%%%%%%%%%%%%%%%%%%%

\section{Current data coverage}
\label{sec:coverage}

The most extensive data coverage for inclusive electron scattering is currently available for the carbon nucleus~\cite{Benhar:2006er}.  Figure~\ref{fig:coverage} shows the $(\omega,Q^2)$ kinematic region covered by these data, compiled from Refs.~\cite{Whitney:1974hr,Barreau:1983ht,O'Connell:1987ag,Bagdasaryan:1988hp,Baran:1988tw,Sealock:1989nx,Day:1993md,Arrington:1995hs,Arrington:1998hz,Arrington:1998ps,Fomin:2008iq,Fomin:2010ei,Dai:2018xhi}. The gray-scale heat map in the background represents the expected event distribution in the DUNE near detector, reproduced from Fig.~\ref{fig:DUNE_distribution}. Each colored curve represents a single dataset, taken at a fixed electron-beam energy and scattering angle.  

Figure~\ref{fig:coverage} demonstrates that---even at the inclusive level and for carbon---there is poor data coverage where the DUNE event density is the highest. As we will see below, in Fig.~\ref{fig:channels}, much of this region is dominated by resonance-excitation and DIS processes, where hadronic physics is highly complex. For improving generator models, it is essential to have not only the inclusive cross sections, but also exclusive measurements that record multiple final-state hadrons. Such measurements are at present not available. 

It is important to note that, even in the context of the inclusive cross sections, the phase space is three-dimensional. That is, three independent kinematic variables are required to fully specify the kinematics: in addition to $\omega$ and $Q^2$, a third variable---such as $\theta_e$ (or beam energy)---needs to be given, in order to calculate the inclusive cross sections. For a~point in the $(\omega, Q^2)$ space, a~good agreement between a cross section estimate and experimental data for some $\theta_e$ does not guarantee that the same is true for significantly different values of~$\theta_e$. The same applies to beam energy. Therefore, for the purpose of long-baseline neutrino program, currently available data for inclusive cross sections cover even smaller fraction of the relevant kinematics than Fig.~\ref{fig:coverage} may suggest at first glance.  Measurements of the cross sections for different scattering angles or beam energies over the same points in the $(\omega, Q^2)$ space are necessary to perform an extensive validation of our description of electroweak interactions with nucleons and atomic nuclei.

Last, we would like to acknowledge that in addition to the cross sections reported in Refs.~\cite{Whitney:1974hr,Barreau:1983ht,O'Connell:1987ag,Bagdasaryan:1988hp,Baran:1988tw,Sealock:1989nx,Day:1993md,Arrington:1995hs,Arrington:1998hz,Arrington:1998ps,Fomin:2008iq,Fomin:2010ei,Dai:2018xhi}, the $F_2(x,Q^2)$ structure functions for deuteron and carbon can currently be validated against the  CLAS measurements performed over a broad kinematics~\cite{Osipenko:2005rh,Osipenko:2010sb}. The LDMX results will complement these inclusive data, as well as provide information on (semi)exclusive cross sections.

%%%%%%%%%%%%%%%%%%%%%%%%%%%%%%%%%%%%%%%%%%%%%%%%%%%%%%%%%%%%%%%%%%%%%%%%%%%%%%
%%%%%%%%%%%%%%%%%%%%%%%%%%%%%%%%%%%%%%%%%%%%%%%%%%%%%%%%%%%%%%%%%%%%%%%%%%%%%%

\section{Inclusive electron distributions}
\label{sec:app_incl}

Figure~\ref{fig:elomega-suppl} shows simulations of the inclusive $e$-Ti cross section for additional scattering angles $\theta_e$, extending the results of Fig.~\ref{fig:elomega}.  We see that there is general disagreement between \genie and \gibuu predictions, at all values of $\theta_e$. Note that due to the trigger selection, $\omega > 1$~GeV, and the transverse momentum cut, $p_T > 200$~MeV, only the events corresponding to the final electron energy $E'> p_T/\sin\theta$ contribute to the distributions presented here.

%%%%%%%%%%%%%%%%%%%%%%%%%%%%%%%%%%%%%%%%%%%%%%%%%%%%%%%%%%%%%%%%%%%%%%%%%%%%%%
%%%%%%%%%%%%%%%%%%%%%%%%%%%%%%%%%%%%%%%%%%%%%%%%%%%%%%%%%%%%%%%%%%%%%%%%%%%%%%

\section{DUNE event distributions}
\label{sec:DUNE_distri}

In Fig.~\ref{fig:channels}, we break down the DUNE event sample simulated with \gibuu according to the individual channels modeled by the generator: quasielastic (QE), meson-exchange current (MEC), resonance production (RES), and deep-inelastic scattering (DIS).  The color scale is consistent in all four panels, i.e., the same color indicates the same event density.  The blue and green lines show constant values of electron-scattering angles $\theta_e$ and transverse momenta $p_T$.

When $Q^2 \simeq 2M\omega$, $M$ being the nucleon mass, the main mechanism of interaction is quasielastic scattering, $\nu_\mu +n\rightarrow \mu^{-} +p$, on individual nucleons inside the nucleus.  Accordingly, we see a linear shape in the top left panel in Fig.~\ref{fig:channels}.  At $Q^2 \simeq 2M\omega + M^2_{\text{res}}-M^2$, the energy transferred to the struck nucleon $N$ is sufficient to excite a baryon resonance state $B_{\text{res}}$ with mass $M_{\text{res}}$, i.e., $\nu_\mu +N\rightarrow \mu^{-} +B_{\text{res}}$.  When the energy transfer increases further, production of higher hadronic resonances gradually transitions to the DIS regime, in which interactions are treated at the quark level.  According to \gibuu, DUNE near detector events are dominated by DIS (39\%) events, closely followed by QE (25\%) and resonance production (24\%) events. 

It should be kept in mind that, physically, the boundary between RES and DIS events---as well as between QE and MEC events---is vague, and is a matter of convention in a given generator. 

%%%%%%%%%%%%%%%%%%%%%%%%%%%%%%%%%%%%%%%%%%%%%%%%%%%%%%%%%%%%%%%%%%%%%%%%%%%%%%
%%%%%%%%%%%%%%%%%%%%%%%%%%%%%%%%%%%%%%%%%%%%%%%%%%%%%%%%%%%%%%%%%%%%%%%%%%%%%%
 
% \bibliographystyle{apsrev4-1}
% \bibliography{references}

\end{document}